\documentclass[aps,prd,amsmath,amssymb,floatfix,nofootinbib,preprint,tightenlines]{revtex4-1}
\usepackage{amsfonts}
\usepackage[pdftex]{graphicx}

\newcommand{\cA}{{\mathcal A}}
\newcommand{\cAb}{{\overline{\mathcal A}}}
\newcommand{\cD}{{\mathcal D}}
\newcommand{\cDb}{{\overline{\mathcal D}}}
\newcommand{\cF}{{\mathcal F}}
\newcommand{\cFb}{{\overline{\mathcal F}}}
\newcommand{\cL}{{\mathcal L}}
\newcommand{\cO}{{\mathcal O}}
\newcommand{\cP}{{\mathcal P}}
\newcommand{\cPtwiddle}{\ensuremath{\widetilde{\mathcal P}} }
\newcommand{\cQ}{{\mathcal Q}}
\newcommand{\cU}{{\mathcal U}}
\newcommand{\cN}{{\mathcal N}}
\newcommand{\cUb}{{\overline{\mathcal U}}}

\newcommand{\tr}{{\rm Tr\;}}
\newcommand{\Tr}{{\rm Tr\;}}

\newcommand{\vn}{ {\bf n} }
\newcommand{\KD}{{K\"{a}hler--Dirac }}
\newcommand{\hatbmu}{\widehat{\boldsymbol{\mu}}}
\newcommand{\hatbe}{\widehat{\boldsymbol{e}}}

\newcommand{\bea}{\begin{align}}
\newcommand{\eea}{\end{align}}
\newcommand{\beq}{\begin{equation}}
\newcommand{\eeq}{\end{equation}}
\newcommand{\nn}{\nonumber }
\newcommand{\re}{\mathrm{Re}}
\newcommand{\im}{\mathrm{Im}}

\def\figheight{7 cm}
\newcommand{\Cbb}{{\mathbb C}}
\newcommand{\Ibb}{{\mathbb I}}
\newcommand{\Rbb}{{\mathbb R}}
\newcommand{\gltwo}{\ensuremath{\mathfrak{gl}(2, \Cbb)} }
\newcommand{\glN}{\ensuremath{\mathfrak{gl}(N, \Cbb)} }
\newcommand{\sltwo}{\ensuremath{\mathfrak{sl}(2, \Cbb)} }
\newcommand{\slN}{\ensuremath{\mathfrak{sl}(N, \Cbb)} }

\renewcommand{\bar}[1]{\ensuremath{\overline #1 } }
\newcommand{\eq}[1]{Eq.~(\ref{#1})}
\newcommand{\fig}[1]{Fig.~\ref{#1}}

\usepackage{color}
\usepackage[pdftex, pdfstartview={FitH}, pdfnewwindow=true, colorlinks=true, citecolor=blue, filecolor=blue, linkcolor=blue, urlcolor=blue, pdfpagemode=UseNone]{hyperref}

\newcommand{\vev}[1]{\ensuremath{\left\langle #1 \right\rangle} }
\newcommand{\pf}{\ensuremath{\mbox{pf}\,} }

\begin{document}
\title{$\cN = 4$ supersymmetry on a space-time lattice}

\author{Simon Catterall}
\affiliation{Department of Physics, Syracuse University, Syracuse, NY 13244, USA}

\author{Poul H. Damgaard}
\affiliation{Niels Bohr International Academy and Discovery Center, Niels Bohr Institute, University of Copenhagen, Blegdamsvej 17, DK-2100 Copenhagen, Denmark}

\author{Thomas DeGrand}
\affiliation{Department of Physics, University of Colorado, Boulder, CO 80309, USA}

\author{Joel Giedt}
\affiliation{Department of Physics, Applied Physics and Astronomy, Rensselaer Polytechnic Institute, 110 8th Street, Troy NY 12065, USA \bigskip}

\author{David Schaich}
\affiliation{Department of Physics, Syracuse University, Syracuse, NY 13244, USA}

\date{29 August 2014 \\ \bigskip}

\begin{abstract}
Maximally supersymmetric Yang--Mills theory in four dimensions can be formulated on a space-time lattice while exactly preserving a single supersymmetry.
Here we explore in detail this lattice theory, paying particular attention to its strongly coupled regime.
Targeting a theory with gauge group SU($N$), the lattice formulation is naturally described in terms of gauge group U($N$).
Although the U(1) degrees of freedom decouple in the continuum limit we show that these degrees of freedom lead to unwanted lattice artifacts at strong coupling.
We demonstrate that these lattice artifacts can be removed, leaving behind a lattice formulation based on the SU($N$) gauge group with the expected apparently conformal behavior at both weak and strong coupling.
\end{abstract}

\maketitle

\section{Introduction}
After close to thirty years of attempts to formulate lattice gauge theories in a manner that would preserve supersymmetry, developments in the last decade led to the construction of lattice versions of a number of interesting supersymmetric theories.
The solution involves no fine tuning in some cases and manageable fine tuning in others.
See Ref.~\cite{Catterall:2009it} for a comprehensive review.
In this work we employ the resulting lattice formulation of four-dimensional maximally supersymmetric $\cN = 4$ Yang--Mills theory to initiate large-scale numerical studies of this system.
In addition to reporting these first numerical results, we provide a detailed review of the lattice construction itself, which possesses several unusual features that may not yet be common knowledge.

At the most basic level, the key observation is that only supercharges that transform as Lorentz scalars can be exactly preserved on a lattice.
These conserved charges anti-commute and also individually square to zero, much like BRST charges.
As we will briefly review in the next section, there exists a very direct link between lattice supersymmetry and BRST symmetry, through a formulation that builds on topological twisting of supersymmetric gauge theories.

In four dimensions the only supersymmetric gauge theory that can be formulated on a space-time lattice in this manner is $\cN = 4$ supersymmetric Yang--Mills (SYM) theory, which has sixteen supersymmetry charges.
On the lattice only one of these charges will be exactly preserved, with the other fifteen broken by lattice artifacts and recovered only in the continuum limit.
In the continuum limit, these sixteen fermionic degrees of freedom give rise to four Majorana (or two Dirac) fermions, in perfect balance with the bosonic degrees of freedom.
While the lattice fermion action that we will present looks quite different from more familiar approaches to lattice fermions, in the free-field limit it can be mapped to a theory of reduced staggered fermions~\cite{Banks:1982iq}.

The lattice construction has several other features that ensure a symmetric treatment of fermions and bosons.
Of course, all fields transform in the adjoint representation.
However, in the usual formulation gauge fields are identified with links on the lattice, while both scalars and fermions are taken to live on sites, a distinction incompatible with supersymmetry.
In addition, the lattice gauge link variables are usually taken to be elements of the gauge group, while the scalar and fermion fields are in the corresponding Lie algebra (as are the gauge potentials in the continuum).

The second issue is easier to address: we simply keep the gauge links in the algebra of the gauge group, on equal footing with the scalar and fermion fields.
A similar approach has been explored in the case of pure (non-supersymmetric) Yang--Mills theory~\cite{Palumbo:1990kh, Becchi:1992um, Palumbo:2001br}.
These constructions feature a flat measure in the partition function, and are most naturally applicable to U($N$) gauge groups.
In this formulation the emergence of an appropriate vacuum, around which the continuum limit can be taken, results from a scalar mode achieving a vacuum expectation value, an interesting dynamical mechanism quite different from the usual Wilson formulation of lattice gauge theory.

Since the gauge fields must necessarily be associated with links to retain gauge invariance, supersymmetry dictates that the fermions should also be located on links.
Such a geometrical prescription can be implemented by working with the \KD equation, as opposed to the usual Dirac equation~\cite{Rabin:1981qj}.
The former describes precisely four degenerate Majorana fermions in the continuum limit, in terms of Grassmann-valued $p$-form fields that may be associated with links in a lattice.
In four dimensions only the theory with $\cN = 4$ supersymmetry possesses the fermion degeneracy required by the \KD equation.
More formally, $p$-forms and the \KD equation arise when the fields of the theory are decomposed under the diagonal subgroup of the SO(4) Euclidean Lorentz symmetry and an SO(4)$_R$ subgroup of the R symmetry of the theory~\cite{Unsal:2006qp}.
This process is the topological twisting procedure mentioned above, and in flat space just corresponds to a change of variables in the theory.
In this way, spin-$1/2$ fields are also distributed along links, on equal footing with the gauge fields.

A similar construction works for the scalar fields, which decompose as an SO(4) vector and two scalars after twisting.
The vector combines with the lattice gauge field to form a complexified link field, i.e., each link is an element of $\glN$.
It is natural to associate the remaining two scalars with a fifth complexified link field, which can be incorporated into the theory by working on the $A_4^*$ lattice whose basis vectors consist of five links symmetrically spanning four space-time dimensions~\cite{Kaplan:2005ta}.
The resulting five-component complexified gauge field decomposes into two irreducible representations under the $S_5$ point group symmetry of the $A_4^*$ lattice: a complex singlet and a complex four-component representation, as required to target $\cN = 4$ SYM in the continuum limit~\cite{Unsal:2006qp}.
To extract physical observables, one simply ``un-twists'' and measures expectation values of combinations of fields that have the appropriate transformation properties under the Euclidean Lorentz group.

Because each link is an element of the algebra $\glN$, the twisted lattice formulation summarized above naturally describes the gauge group U($N$), the maximal compact subgroup of GL($N, \Cbb$).
Although the U(1) gauge degrees of freedom decouple in the continuum limit, where U($N$) $=$ SU($N$)$\otimes$U(1), they introduce unwanted lattice artifacts at strong coupling.
A new development in the work we present below is a mechanism to remove these lattice artifacts, leaving behind a lattice formulation based on SU($N$).
Specifically, we add a deformation to the lattice action, which softly breaks the nilpotent Lorentz-scalar supersymmetry.
Additional soft supersymmetry breaking is introduced by another deformation included to regulate flat directions in numerical calculations.

Our approach does not exhaust the possibilities for numerically studying $\cN = 4$ SYM with gauge group SU($N$).
An intriguing alternative method employs the equivalence between this theory in the large-$N$ limit on the manifold $\Rbb\times S^3$ and a certain limit of supersymmetric large-$N$ quantum mechanics~\cite{Ishii:2008ib, Ishiki:2008te, Ishiki:2009sg, Honda:2011qk, Honda:2013nfa, Hanada:2013rga}.
This treats all 16 supercharges on an equal footing, but none of them are exactly preserved at a finite ultraviolet cut-off.
There is also a proposal~\cite{Hanada:2010kt} to preserve two supersymmetries by working with two commutative and two noncommutative dimensions.

This paper is organized as follows: in the next two sections we review in more detail the twisted lattice formulation of $\cN = 4$ SYM briefly summarized above, then spend two sections presenting some first results from our large-scale numerical computations.
We begin by considering the continuum action in terms of the twisted fields, and show how this system can be directly transcribed to produce the lattice theory.
In Section~\ref{sec:A4star} we review the $A_4^*$ lattice itself, and the role that its structure plays in comparisons of lattice results with continuum expectations.
We then discuss how the continuum limit of the theory is recovered from lattice calculations.
Section~\ref{sec:UvsSU} first shows that the U(1) sector of the U($N$) gauge group induces a transition to a strongly-coupled lattice phase with no analog in the continuum theory.
We then demonstrate that this problem may be cured by adding to the lattice action a new term that suppresses the U(1) gauge degrees of freedom, leaving behind a lattice formulation based on gauge group SU($N$).

Using this new lattice action, we have initiated large-scale numerical studies of $\cN = 4$ SYM.
The first results that we present in Section~\ref{sec:tests} provide evidence that our lattice theory simulates SYM to a good approximation, despite the deformations necessary to carry out numerical calculations.
In addition, we show that direct measurements of the pfaffian indicate no sign problem, and take a first look at the large-$N$ limit, where we observe $1 / N^2$ suppression of supersymmetry-breaking effects and no change in the pfaffian phase.
Finally, in Section~\ref{sec:potential} we study the static potential, which exhibits Coulombic behavior at both weak and strong coupling.
We consider potentials based on three different types of Wilson loops, and find that their Coulomb coefficients have the expected relative magnitudes.
We conclude with a look forward at further investigations that we will soon present in future publications.

\section{Connecting the continuum and lattice theories}
In the continuum, $\cN = 4$ SYM can be twisted in different ways to form topological field theories.
The known lattice construction closely mimics the twisting introduced by Marcus~\cite{Marcus:1995mq}, which is sometimes called the GL-twist due to its important role in the Geometric Langlands program~\cite{Kapustin:2006pk}.
In this section we summarize the twisted theory and its transcription to the $A_4^*$ lattice, some features of which will play important roles in the numerical results presented in Sections~\ref{sec:tests} and \ref{sec:potential}.
Although most of the information in this section has already appeared in Refs.~\cite{Kaplan:2005ta, Unsal:2006qp, Catterall:2009it, Catterall:2012yq} among others (the new results in Section~\ref{sec:continuum} will be derived in a separate future publication~\cite{Catterall:2014mha}), we suspect that much of it is not yet common knowledge, which motivates this review.

\subsection{\label{sec:basics}From topological twisting to the lattice action}
The central idea of the GL-twist is to form the complex combination
\beq
  \label{4dgauge}
  \cA_\mu \equiv A_{\mu} + iB_{\mu},
\eeq
where $A_{\mu}$ are the usual four-dimensional gauge potentials and $B_{\mu}$ is a vector formed out of four of the six adjoint scalars $\Phi_{IJ}$ of the $\cN = 4$ theory.
The two remaining scalars remain singlets after twisting.
The four Majorana fermions ($\Psi^I$ and CPT-conjugate partners $\Psi^c_I$) are regrouped into an anti-symmetric tensor $\chi_{\mu\nu}$, two vectors $\psi_\mu$ and $\bar{\psi}_{\mu}$ and two scalar components $\eta$ and $\bar{\eta}$, altogether 16 single components.

It turns out to be more natural to work in terms of five-component objects, extending the complex gauge combination to
\beq
  \label{5dgauge}
  \cA_a \equiv A_a + iB_a,
\eeq
where the roman index ``$a$'' runs from $1, \cdots, 5$.
We assign the two singlet scalars to the new fifth component $\cA_5$.
Similarly, the 16 fermionic fields can be regrouped into the multiplet $\chi_{ab}, \psi_a, \eta$, with $\chi_{ab}$ still anti-symmetric.
We can then introduce complexified field strengths,
\begin{align}
  \cF_{ab} & \equiv [\cD_a, \cD_b] &
  \cFb_{ab} & \equiv [\cDb_a, \cDb_b],
\end{align}
where the corresponding complexified covariant derivatives read
\begin{align}
  \cD_a & = \partial_a + \cA_a &
  \cDb_a & = \partial_a + \cAb_a.
\end{align}
One scalar supersymmetry charge $\cQ$ takes on the meaning of a BRST charge after the twisting.
In the notation introduced above, it acts as follows:
\begin{align}
  & \cQ\; \cA_a = \psi_a         & & \cQ\; \psi_a = 0                       \cr
  & \cQ\; \chi_{ab} = -\cFb_{ab} & & \cQ\; \cAb_a = 0 \label{BRSTsymmetry}  \\
  & \cQ\; \eta = d               & & \cQ\; d = 0,                           \nn
\end{align}
where $d$ is a bosonic auxiliary field with equation of motion $d = \left[\cDb_a, \cD_a\right]$ (repeated indices summed).
The other fifteen supersymmetry charges are twisted into a vector $\cQ_a$ and anti-symmetric tensor $\cQ_{ab}$.

Except for a topological $\cQ$-closed term,
\beq
  \label{closed}
  S_{\rm cl} = -\frac{1}{8g^2} \int \Tr \epsilon_{mnpqr} \chi_{qr} \cDb_p \chi_{mn},
\eeq
the full $\cN = 4$ action can be written as the BRST gauge fixing of arbitrary field deformations:
\beq
  \label{4daction}
  S = \frac{1}{2g^2} \cQ \int \mbox{Tr}\left[\chi_{ab}\cF_{ab} + \eta [ \cDb_a,\cD_a ] - \frac{1}{2}\eta d\right] + S_{\rm cl}
\eeq
where $\cQ\; S_{\rm cl} = 0$ is guaranteed by the Bianchi identity.
Since we continue to work in four space-time dimensions, the symmetric constraint $\sum_a \partial_a = 0$ provides the proper number of independent differentiations.
Equivalently one can obtain the $\cN = 4$ theory by a naive dimensional reduction of the five-dimensional theory.

As explained in detail in Ref.~\cite{Catterall:2007kn}, this twisted formulation leads naturally to a lattice construction of the theory.
In fact, there is a very direct and geometric prescription for how to map continuum variables (covariant derivatives and tensor fields of arbitrary rank) to those of the lattice~\cite{Catterall:2007kn, Damgaard:2008pa}.
In this particular case, the lattice inherits the five-component language, and is most naturally represented as the $A_4^*$ lattice with manifest $S_5$ point group symmetry in four space-time dimensions.
The basis vectors of the $A_4^*$ lattice link the center of an equilateral 4-simplex to each of its five vertices.
This is the analog of the triangular lattice in two dimensions, and we will review its properties in the next subsection.

In terms of the complex link variables $\cU_a(\vn)$, and the finite difference operators
\begin{align}
  \cD^{(+)}_a f_b(\vn) & = \cU_a(\vn)f_b(\vn + \hatbmu_a) - f_b(\vn)\cU_a(\vn+ \hatbmu_b)                                       \cr
  \cDb^{(-)}_a f_a(\vn) & = f_a(\vn)\cUb_a(\vn) - \cUb_a(\vn - \hatbmu_a)f_a(\vn - \hatbmu_a)                                   \\
  \cDb_c^{(-)} f_{ab}(\vn + \hatbmu_c) & = f_{ab}(\vn + \hatbmu_c) \cUb_c(\vn + \hatbmu_a + \hatbmu_b) - \cUb_c(\vn)f_{ab}(\vn) \nn
\end{align}
from Refs.~\cite{Catterall:2007kn, Damgaard:2008pa}, the lattice action can be written down by transcribing the continuum action:
\begin{align}
  \label{SlatQ}
  S_0 & = \frac{N}{2\lambda_{\rm lat}} \sum_{\vn} \Tr \cQ \left(\chi_{ab}(\vn)\cD_a^{(+)}\cU_b(\vn) + \eta(\vn) \cDb_a^{(-)}\cU_a(\vn) - \frac{1}{2}\eta(\vn) d(\vn) \right) + S_{\rm cl} \\
  S_{\rm cl} & = -\frac{N}{8\lambda_{\rm lat}} \sum_{\vn} \Tr \epsilon_{abcde} \chi_{de}(\vn + \hatbmu_a + \hatbmu_b + \hatbmu_c) \cDb^{(-)}_{c} \chi_{ab}(\vn + \hatbmu_c).
\end{align}
Here $\lambda_{\rm lat} = g_{\rm lat}^2 N$ differs from the continuum 't~Hooft coupling by a normalization factor of $1 / \sqrt{5}$, which we derive in the next subsection.
On the lattice $S_{\rm cl}$ is $\cQ$-closed on account of a lattice analog of the continuum Bianchi identity~\cite{Catterall:2007kn},
\begin{equation}
  \epsilon_{abcde} \cDb^{(-)}_c \cFb_{ab}(\vn + \hatbmu_c) = 0.
\end{equation}
We see that as a ``topological lattice theory'' the action (\ref{SlatQ}) gauge fixes on our lattice analog of complexified flat connections,
\begin{equation}
  \cD_a^{(+)}\cU_b(\vn) = \cU_a(\vn) \cU_b(\vn + \hatbmu_a) - \cU_b(\vn) \cU_a(\vn + \hatbmu_b) = 0,
\end{equation}
but we stress that this pseudo-topological interpretation of our lattice action is irrelevant here.
The physical observables are those of the untwisted theory to which we can always map back.

Expanding the action (\ref{SlatQ}) using the lattice analog of \eq{BRSTsymmetry},
\begin{align}
  & \cQ\; \cU_a = \psi_a          & & \cQ\; \psi_a = 0 \nn \\
  & \cQ\; \chi_{ab} = -\cFb_{ab}  & & \cQ\; \cUb_a = 0 \label{BRSTlatticesymmetry} \\
  & \cQ\; \eta = d                & & \cQ\; d = 0, \nn
\end{align}
and integrating out the auxiliary field $d$ one obtains
\begin{equation}
  \label{eq:lat_act}
  \begin{split}
    S_0 & = \frac{N}{2\lambda_{\rm lat}} \sum_{\vn} \mbox{Tr}\left[-\cFb_{ab}(\vn) \cF_{ab}(\vn) + \frac{1}{2}\left(\cDb_a^{(-)}\cU_a(\vn)\right)^2 \right. \\
        & \qquad\qquad\qquad\qquad\qquad\qquad\qquad \left. - \chi_{ab}(\vn) \cD^{(+)}_{[a}\psi_{b]}(\vn) - \eta(\vn) \cDb^{(-)}_a\psi_a(\vn)\right] + S_{\rm cl}.
  \end{split}
\end{equation}
As discussed in Ref.~\cite{Catterall:2012yq}, to stabilize numerical computations we regulate the flat directions by including in the lattice action a potential
\begin{equation}
  \label{eq:mass}
  S = S_0 + \frac{N}{2\lambda_{\rm lat}} \mu^2 \sum_{\vn,\ a} \left(\frac{1}{N}\mbox{Tr}\left[\cUb_a(\vn) \cU_a(\vn)\right] - 1\right)^2,
\end{equation}
with $\mu$ a tunable ``bosonic mass'' parameter.
Non-zero $\mu$ softly breaks supersymmetry, an issue we explore in Section~\ref{sec:tests}.
The full action is invariant under lattice gauge transformations,
\begin{align}
  \cU_a(\vn) & \to G(\vn)\cU_a(\vn)G^{\dag}(\vn+\hatbmu_a)                   &
  \psi_a(\vn) & \to G(\vn)\psi_a(\vn)G^{\dag}(\vn+\hatbmu_a)                 \\
  \chi_{ab}(\vn) & \to G(\vn+\hatbmu_a+\hatbmu_b)\chi_{ab}(\vn)G^{\dag}(\vn) &
  \eta(\vn) & \to G(\vn)\eta(\vn)G^{\dag}(\vn),                              \nn
\end{align}
where $G \in$ U($N$).
These transformation rules are as expected for lattice variables in the adjoint representation.
What is new here is that the gauge links
\beq
  \cU_a(\vn) = \sum_{C = 1}^{N^2} T^C\cU^C_a(\vn)
\eeq
are expanded in generators of the $\mathfrak u(N)$ algebra with complex coefficients.
(We use anti-hermitian generators normalized by $\Tr(T^AT^B) = -\delta^{AB}$.)
These links are therefore elements of the algebra $\glN$.

To obtain the correct naive continuum limit, the complexified gauge links must have the expansion $\cU_a(x) = \Ibb + \cA_a(x) + \dots$ in some appropriate gauge.
This requirement can be satisfied by arranging for the imaginary part of the U(1) component of $\cU_a$ to take on a vacuum expectation value.
This vacuum expectation value sets the lattice scale that allows the theory to define a derivative in the continuum limit.
The bosonic mass term in \eq{eq:mass} stabilizes just such a vacuum state.
We discuss the continuum limit in more detail below, following a review of the $A_4^*$ structure underlying our lattice system.

\subsection{\label{sec:A4star}The underlying $A_4^*$ lattice structure}
Dealing with the $A_4^*$ lattice presents some conceptual issues which, while straightforward, require a little discussion.
We define the lattice variables on an abstract hypercubic lattice that includes an additional body diagonal link:
\begin{align}
  \hatbmu_1 & = (1, 0, 0, 0)                  \cr
  \hatbmu_2 & = (0, 1, 0, 0)                  \cr
  \hatbmu_3 & = (0, 0, 1, 0) \label{eq:dirs}  \\
  \hatbmu_4 & = (0, 0, 0, 1)                  \cr
  \hatbmu_5 & = (-1, -1, -1, -1).             \nn
\end{align}
Relative to this abstract basis a field can be given integer coordinates $\vn = (n_1, n_2, n_3, n_4)$.
A specific basis for the $A_4^*$ lattice takes the form of five lattice vectors~\cite{Kaplan:2005ta}
\begin{align}
  \hatbe_1 & = \left(\frac{1}{\sqrt{2}}, \frac{1}{\sqrt{6}}, \frac{1}{\sqrt{12}}, \frac{1}{\sqrt{20}}\right)  \cr
  \hatbe_2 & = \left(-\frac{1}{\sqrt{2}}, \frac{1}{\sqrt{6}}, \frac{1}{\sqrt{12}}, \frac{1}{\sqrt{20}}\right) \cr
  \hatbe_3 & = \left(0, -\frac{2}{\sqrt{6}}, \frac{1}{\sqrt{12}}, \frac{1}{\sqrt{20}}\right) \label{eq:basis} \\
  \hatbe_4 & = \left(0, 0, -\frac{3}{\sqrt{12}}, \frac{1}{\sqrt{20}}\right)                                   \cr
  \hatbe_5 & = \left(0, 0, 0, -\frac{4}{\sqrt{20}}\right).                                                    \nn
\end{align}
The basis vectors satisfy the relations
\begin{align}
  & \sum_{m = 1}^5 \hatbe_m = 0                                       &
  & \hatbe_m \cdot \hatbe_n = \left(\delta_{mn} - \frac{1}{5}\right)  &
  & \sum_{m = 1}^5 (\hatbe_m)_{\mu} (\hatbe_m)_{\nu} = \delta_{\mu\nu}
\end{align}
for $\mu, \nu = 1, \cdots, 4$.
The physical location of a lattice field is simply $\displaystyle r = \sum_{\nu = 1}^4 \hatbe_{\nu} n_{\nu}$.

Using these basis vectors (\ref{eq:basis}) one can construct a $5\times 5$ orthogonal matrix $P$ where
\begin{align}
  \label{eq:P}
  P_{\mu a} & = (\hatbe_a)_{\mu} &
  P_{5 a} & = \frac{1}{\sqrt{5}}
\end{align}
for $\mu = 1, \cdots, 4$ and $a = 1, \cdots, 5$.
As shown in Ref.~\cite{Unsal:2006qp}, this matrix allows us to write down the irreducible representations of the theory under the lattice point group $S_5$.\footnote{Strictly, the representations are under the discrete rotational subgroup $A_5 = S_5 / Z_2$.}
For example, consider the five-component fermion $\psi_a$.
Under $S_5$ this transforms as $\psi_a^{\prime} = O_{ab}\psi_b$ where $O(g)$ is the matrix representation of the group operation $g$.
In general this group action is reducible, which implies that there is a similarity transformation that block diagonalizes $O$.
In our case the two irreducible components have dimensions 4 and 1, where the singlet is just $\frac{1}{\sqrt 5}\sum_a \psi_a$.
Notice that the dimensions of these representations of $S_5$ match precisely those of representations of the continuum SO(4) twisted rotation group.
This makes it clear that indeed in the continuum limit one expects a vector fermion $\psi_{\mu}$ and an additional scalar fermion $\bar{\eta}$ to arise precisely as in the twisted theory.

Indeed the matrix $P$ makes this connection to the continuum explicit: the transformed $\psi_{\mu}^{\prime} = P_{\mu a} \psi_a$ with $\mu = 1, \cdots, 4$ yield the vector, while $\psi_5^{\prime}$ is the scalar.
The story is identical for the complexified gauge links and confirms that the $A_4^*$ lattice theory yields a gauge field $A_{\mu}$, four scalars $B_{\mu}$ and two additional scalars in the continuum limit.
Similar considerations apply to the ten fields in $\chi_{ab}$, which break up into six- and four-dimensional irreducible representations under $S_5$.
The mapping is just given by
\begin{equation}
  \chi_{\mu\nu} = P_{\mu a} P_{\nu b} \chi_{ab}.
\end{equation}
The six-dimensional representation corresponds to restricting the indices $\mu$, $\nu < 5$, while the four-dimensional vector corresponds to the second vector fermion $\chi_{5\mu} = \bar{\psi}_{\mu}$ arising in the continuum twisted theory.
It is quite remarkable that the low-lying representations of the lattice point group symmetry match perfectly the corresponding representations of SO(4) in the continuum theory.
Indeed, using this technology it is straightforward to show that in the naive continuum limit the lattice action given in \eq{SlatQ} correctly yields the continuum twisted SYM theory as written down by Marcus~\cite{Marcus:1995mq}.

This comparison between the lattice and continuum actions requires rescaling the lattice coupling $\lambda_{\rm lat} = g_{\rm lat}^2 N$, because the $A_4^*$ basis vectors in \eq{eq:basis} are not orthogonal.
This issue was encountered in non-supersymmetric lattice studies during the early 1980s~\cite{Drouffe:1983kq, Ardill:1983hn, Celmaster:1984pv}, and is discussed for $\cN = 4$ SYM in Ref.~\cite{Kaplan:2005ta}.
We can quickly understand it by considering a continuum field theory in a $D$-dimensional finite volume $V = L^D$.
Compare this to the lattice theory, for which $V_{\rm lat} = \sum_{\vn} V_0 = N^D V_0$ is defined by summing the volume $V_0$ of the unit cell over lattice sites $\vn$ labelled by integers from 1 to $N$ along each independent direction.
Taking the lattice sites in these independent directions to be separated by a constant distance ``$a$'', then $L = a N$, and $V_{\rm lat}$ only equals $V$ if $V_0 = a^D$, which in turn only occurs for a hypercubic lattice.
For our $A_4^*$ lattice defined by the basis (\ref{eq:basis}), the unit cell is a 4-parallelotope (a generalized parallelepiped) with volume $V_0 = a^4 \det P_{\mu\nu} = a^4 / \sqrt 5$.
For the $A_D^*$ lattice in $D$ dimensions, $V_0 = a^D / \sqrt{D + 1}$ involves the determinant of the first $D\times D$ components of $P$ (\ref{eq:P}).

We can recognize this $V_0 = a^4 / \sqrt 5$ as the jacobian of the transformation between the lattice and continuum space-time coordinates.
Therefore the action
\begin{equation}
  \label{eq:g_norm}
  S = \frac{1}{g^2} \int_V d^Dx \cL(x) = \frac{V_0}{g^2} \sum_{\vn} \cL(\vn) \equiv \frac{a^4}{g_{\rm lat}^2} \sum_{\vn} \cL(\vn),
\end{equation}
identifying the lattice coupling $g_{\rm lat}^2 = g^2 \sqrt 5$.
That is, numerical calculations using a given value of $\lambda_{\rm lat} = g_{\rm lat}^2 N$ correspond to a weaker continuum 't~Hooft coupling, by a factor of $1 / \sqrt{5}$.
This normalization factor affects comparisons of lattice results with continuum expectations, as we will see for the static potential in Section~\ref{sec:potential}.

\subsection{\label{sec:continuum}The continuum limit}
Because lattice studies of $\cN = 4$ SYM are still in their early stages, it remains an important task to verify that the continuum limit of the lattice system correctly recovers all the symmetries of continuum theory: the SO(6) Euclidean conformal symmetry and the SO(6)$_R$ $\simeq$ SU(4)$_R$ R symmetry
The GL-twist provides fields that transform as multiplets under the twisted SO(4) rotation group, the diagonal subgroup of SO(4) Euclidean Lorentz symmetry and an SO(4)$_R$ $\subset$ SO(6)$_R$.
On the $A_4^*$ lattice we retain the discrete $S_5$ subgroup of the twisted rotation group discussed in the previous subsection.
The lattice action is also invariant under a U(1)$^4$ center symmetry,
\begin{align}
  \cU_a & \to z_a \cU_a                       &
  \cUb_a & \to z_a^{-1} \cUb_a                &
  \psi_a & \to z_a \psi_a                     &
  \chi_{ab} & \to z_a^{-1} z_b^{-1} \chi_{ab} &
  \eta & \to \eta,
\end{align}
where $z_a = \exp(i\phi_a)$ for $\phi_a \in \Rbb$ subject to the constraint $\phi_1 + \phi_2 + \phi_3 + \phi_4 + \phi_5 = 0$.

While the lattice action in \eq{eq:lat_act} is invariant under the scalar supersymmetry $\cQ$, non-zero $\mu$ in \eq{eq:mass} softly breaks this symmetry.
The other fifteen supersymmetries $\cQ_a$ and $\cQ_{ab}$ are broken by the lattice discretization itself, while the finite lattice volume and non-zero lattice spacing break conformal symmetry.
All of these symmetries must be recovered in the continuum limit.

We obtain the continuum limit by taking $1 / L \to 0$, where $L$ is the linear length of the lattice volume~\cite{Catterall:2012yq}.
Since $\cN = 4$ SYM possesses a line of conformal fixed points, this continuum limit may be taken for any fixed value of the bare 't~Hooft coupling $\lambda_{\rm lat}$.
To recover supersymmetry, we must also tune $\mu \to 0$; like $1 / L$, it is a relevant parameter of the lattice system.
Because the scalar modes stabilized by non-zero $\mu$ are constant in space and time, we should only need to hold $\mu^2 L^D$ fixed in numerical computations, which would allow us to systematically decrease $\mu$ in tandem with $1 / L$.

As we approach the continuum limit, the long-distance effective action of the lattice theory takes the form
\begin{equation}
  \label{eq:S_eff}
  \begin{split}
    S_{eff} & = \cQ \; \tr \left\{ \alpha_1 \chi_{ab} \cF_{ab} + \alpha_2 \eta \cDb_a^{(-)} \cU_a - \frac{\alpha_3}{2} \eta d \right\} - \frac{\alpha_4}{4} \epsilon_{abcde} \tr \chi_{de} \cDb_c^{(-)} \chi_{ab} \\
   & \qquad + \beta \cQ \left\{ \tr \eta \cU_a \cUb_a - \frac{1}{N} \tr \eta \tr \cU_a \cUb_a \right\},
  \end{split}
\end{equation}
up to non-renormalizable terms.
Ref.~\cite{Catterall:2011pd} studied this effective action at weak coupling in perturbation theory, reaching the conclusion that mass terms are not generated at any order, and that the divergent parts of the renormalizations at one loop are universal, and hence do not violate the full supersymmetry.
More recently, a blocking scheme has been developed which preserves the $\cQ$ supersymmetry and the lattice structure~\cite{Catterall:2014mha}.
The effective action above is the most general renormalizable action consistent with the symmetries preserved by this blocking transformation.

Recovery of the full $\cN = 4$ supersymmetry in the continuum limit corresponds to universal $\alpha_i$ and vanishing $\beta$ in \eq{eq:S_eff}.
Recently it has been realized, in Ref.~\cite{Catterall:2013roa}, that discrete R symmetries (subgroups of the continuum SU(4)$_R$ symmetry) enforce the desired conditions $\alpha_i = \alpha \; \forall \; i$ and $\beta = 0$.
In addition it can be seen that the recovery of the Euclidean Lorentz symmetry also follows from these R symmetries, so there is a connection between all of these outcomes.

The remaining question is the extent to which the parameters in \eq{eq:S_eff} must be fine-tuned in order to realize the necessary discrete R symmetries in the lattice system, and thus recover the complete symmetries of continuum $\cN = 4$ SYM as we take the $1 / L \to 0$ continuum limit.
Although five parameters appear in \eq{eq:S_eff}, it is possible to show that even the worst-case scenario requires at most two fine tunings to target the correct continuum theory~\cite{Catterall:2014mha}.
In Section~\ref{sec:other_susies} we discuss newly-developed observables that allow us to monitor the breaking and restoration of the relevant discrete R symmetries on the lattice, and present promising initial results.

\section{\label{sec:UvsSU}U($N$) versus SU($N$) gauge group}
A peculiar feature of the construction described in the previous section is that it is based on gauge group U($N$) rather than the preferred SU($N$).
This is a consequence of combining the gauge and scalar fields into complexified link variables, which are elements of the algebra $\glN$.
The maximal compact subgroup of GL($N, \Cbb$) is U($N$).

In the end, we are only interested in the SU($N$) degrees of freedom.
In the continuum this is not an issue, as U($N$) should be viewed as the product gauge group SU($N$)$\otimes$U(1) and the U(1) sector decouples from observables in the SU($N$) sector.
At non-zero lattice spacing, the decoupling of the compact U(1) degrees of freedom is not automatic.
More importantly, as we shall now see, these compact Abelian gauge degrees of freedom cause our lattice theory to enter a phase totally dominated by lattice artifacts at strong coupling.
This issue was studied in the early 1980s for non-supersymmetric U($N$) gauge theories on both hypercubic and simplicial lattices~\cite{Creutz:1982kf, Drouffe:1983vv}.

\begin{figure*}[btp]
  \centering
  \includegraphics[height=\figheight]{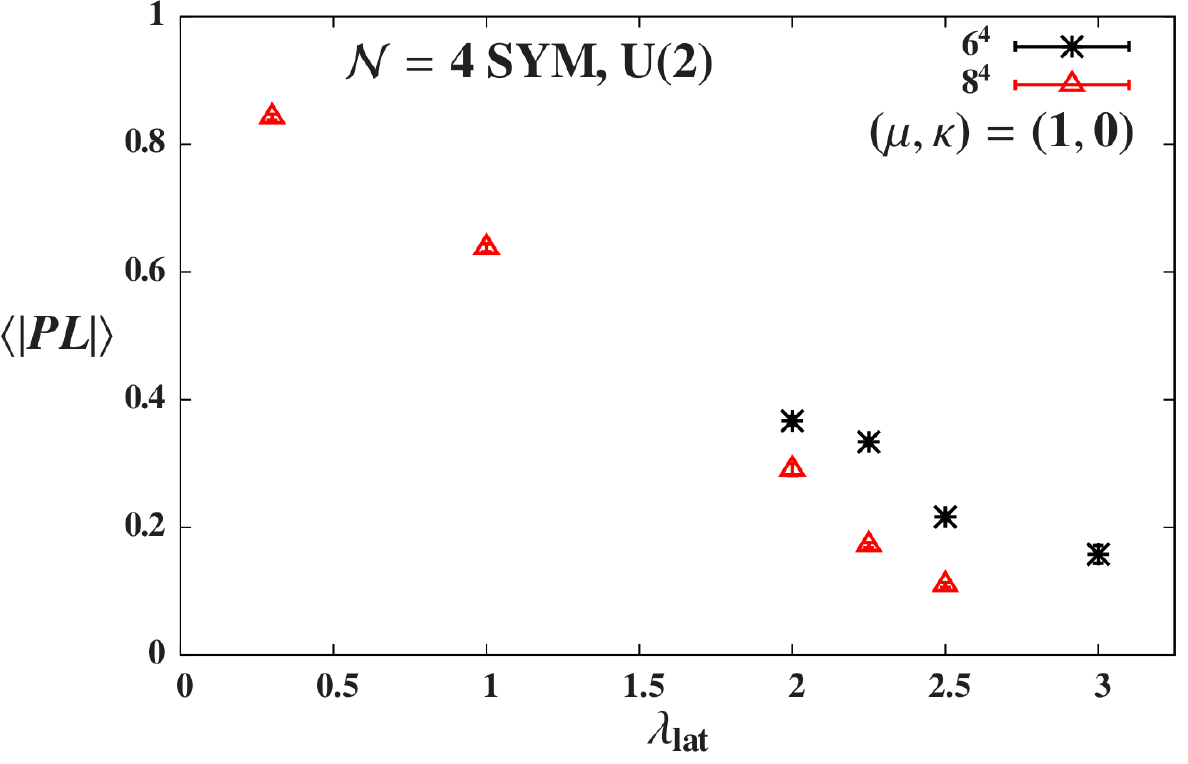}
  \caption{\label{fig:pl_vs_lambda}The absolute value of the Polyakov loop plotted vs.\ the 't~Hooft coupling $\lambda_{\rm lat}$ for fixed $\mu = 1$ indicates that the lattice theory defined by Eqs.~\protect\ref{eq:lat_act} and \protect\ref{eq:mass} transitions into a confining phase at strong coupling, characterized by $\vev{|PL|} \ll 1$.}
\end{figure*}

To illustrate the problem, \fig{fig:pl_vs_lambda} shows numerical results for the absolute value of the Polyakov loop on small $L^4 = 6^4$ and $8^4$ lattices with gauge group U(2), as a function of $\lambda_{\rm lat}$ at fixed $\mu = 1$.
We expect $\cN = 4$ SYM to be deconfined, so that $\vev{|PL|}$ should be approximately unity, as is the case at weak coupling.
As $\lambda_{\rm lat}$ increases, however, the Polyakov loop collapses to a small, $L$-dependent value consistent with a confining theory.
(The lattices considered in \fig{fig:pl_vs_lambda} are too small to show a bona fide transition.)
All numerical results presented in this paper employ anti-periodic (thermal) temporal boundary conditions for the fermions.
The collapse of the Polyakov loop is still present when we use periodic boundary conditions in all four directions.

\begin{figure*}[btp]
  \centering
  \includegraphics[height=\figheight]{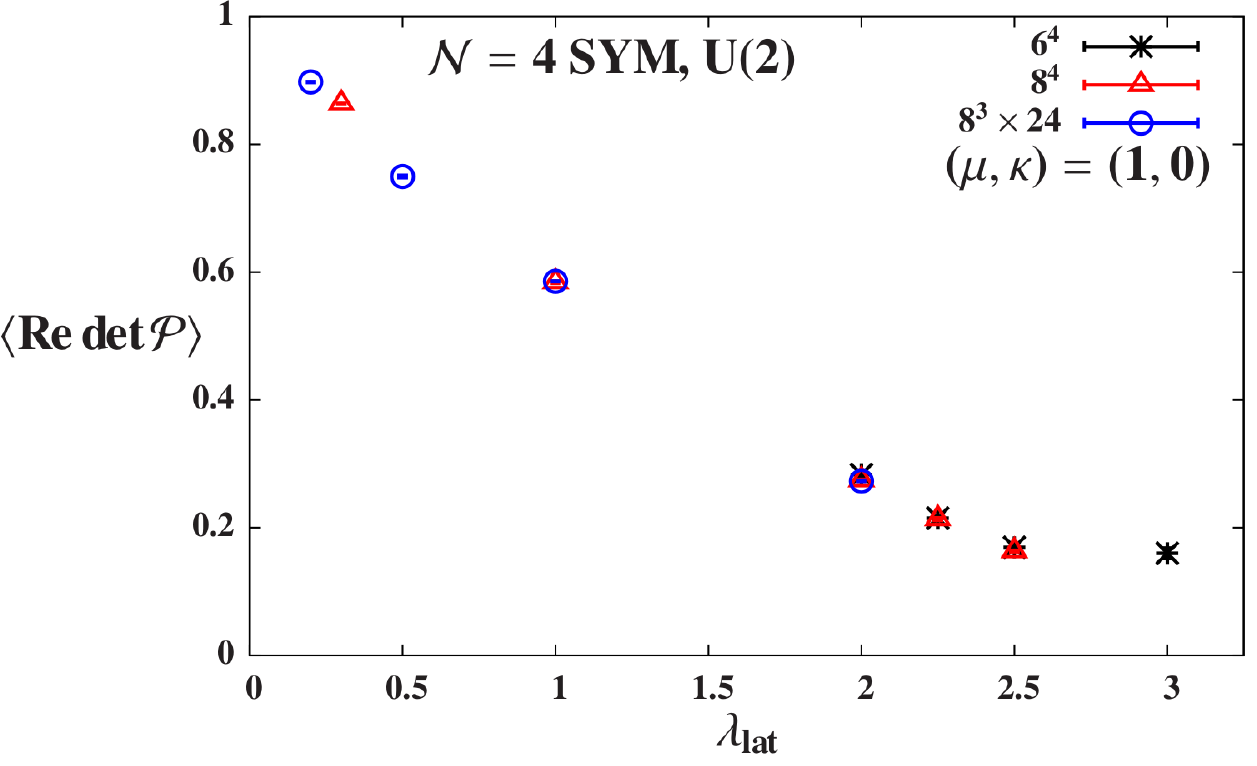}
  \caption{\label{fig:det_vs_lambda}The real part of the plaquette determinant is a gauge-invariant quantity associated with the U(1) sector.  As a function of the 't~Hooft coupling $\lambda_{\rm lat}$ for fixed $\mu = 1$, it shows the same confinement transition at strong coupling as the Polyakov loop in \protect\fig{fig:pl_vs_lambda}, suggesting that this transition is related to the U(1) sector.}
\end{figure*}

The behavior of the Polyakov loop indicates that the lattice theory has a phase inconsistent with our expectations for $\cN = 4$ SYM.
We are able to attribute this behavior to the U(1) sector, which was also the case for lattice transitions observed in the studies of the non-supersymmetric U($N$) gauge theories mentioned above~\cite{Creutz:1982kf, Drouffe:1983vv}.
First consider \fig{fig:det_vs_lambda}, which shows the real part of the determinant of the plaquette.
This plaquette determinant is a gauge-invariant quantity associated with the U(1) sector, and its expectation value decreases sharply around the same $\lambda_{\rm lat}$ values where $\vev{|PL|}$ collapses in \fig{fig:pl_vs_lambda}.
This suggests a connection between the U(1) sector and the strong-coupling lattice phase, which we can make more precise by studying magnetic monopoles in the U(1) sector.


\subsection{Magnetic monopoles in the U(1) sector at strong coupling}
Recall that compact U(1) lattice gauge theory undergoes a phase transition from a Coulombic weak-coupling phase to a confined strong-coupling phase~\cite{DeGrand:1980eq}.
Above we suggested that the behavior of the Polyakov loop in \fig{fig:pl_vs_lambda} and the plaquette determinant in \fig{fig:det_vs_lambda} could result from similar physics in the U(1) sector of our lattice system.
To study this question, here we consider the dual description of the U(1) sector in terms of magnetic variables.

In this dual description, ordinary (``electric'') confinement in the strong-coupling phase corresponds to screened magnetic fields.
In three dimensions the dual variables are point defects (instantons), and the dual description of confinement was famously first given by Polyakov~\cite{Polyakov:1976fu}.
In four dimensions, the dual variables are one-dimensional objects, magnetic world lines, which form closed loops of monopole flux.
The Coulombic weak-coupling phase is associated with a small density of monopole world lines, while in the confined strong-coupling phase the density of monopole world lines is large.

The technique that we employ is a small variation on the methodology of Ref.~\cite{DeGrand:1980eq}.
We identify monopoles or monopole world lines by counting Dirac strings.
To do this we need to associate a U(1) variable with each link $\cU_{\mu}(x)$.
Considering first gauge fields on a hypercubic lattice, the phase of the determinant of the link is
\begin{equation}
  \phi_{\mu}(x) = \tan^{-1} \frac{\im \det \cU_{\mu}(x)}{\re \det \cU_{\mu}(x)},
\end{equation}
which runs between $-\pi$ and $\pi$.
The phase $\Phi_{\mu\nu}$ associated with each plaquette is the sum of the oriented phases on the edges, and so can run from $-4\pi$ to $4\pi$.
This is gauge invariant because of the product rule for determinants.
We define an integer Dirac string number $N_{\mu\nu}$ by $\Phi_{\mu\nu} = \Phi_{\mu\nu}^{\prime} + 2\pi N_{\mu\nu}$, where $-\pi < \Phi_{\mu\nu}^{\prime} \leq \pi$.
The rationale is that $F_{\mu\nu}$ in the U(1) sector, represented by $\Phi_{\mu\nu}^{\prime}$, should have the same range as $\phi_{\mu}$, the U(1) variable on the link.

Now we define a monopole by counting Dirac strings.
This is easiest to visualize in three dimensions where monopoles are point objects, living inside cubes.
We count the number of strings entering or leaving each face of the cube, and their net sum is the charge
\beq
  M = \epsilon_{\mu\nu\lambda} \Delta_{\mu}^{(+)} N_{\nu\lambda},
\eeq
with $\Delta_{\mu}^{(+)} \psi(\vn) = \psi(\vn + \hatbmu) - \psi(\vn)$.
In four dimensions, defects are actually one dimensional objects, monopole world lines.
The local monopole current in direction $\mu$ is
\beq
  \label{worldline}
  M_{\mu} = \epsilon_{\mu\nu\rho\sigma} \Delta_{\nu}^{(+)} N_{\rho\sigma}.
\eeq
It is hard to follow the monopole world lines, but it is easy to check that the sum of all charge (or current in a particular direction) adds up to zero.
The interesting quantity is the total perimeter of monopole strings, which is just the sum of the absolute values of $M_{\mu}$ over all sites and directions.

On the $A_4^*$ lattice, we should consider an object which transforms covariantly under the $S_5$ point group symmetry,
\begin{equation}
  M_{ab} = \epsilon_{abcde}\Delta_c^{(+)} N_{de}.
\end{equation}
To transform this to continuum-like coordinates, and separate the four gauge fields in $\cU_{\mu}$ from the additional scalars in $\cU_5$, we apply the orthogonal matrix $P$ from \eq{eq:P}:
\begin{equation}
  M_{\mu\nu} = P_{\mu a}P_{\nu b} M_{ab} = \epsilon_{abcde} P_{\mu a}P_{\nu b} P_{c \rho}^T P_{d \sigma}^T P_{e \lambda}^T \Delta_{\rho}^{(+)} N_{\sigma\lambda} = (\det P) \epsilon_{\mu\nu\rho\sigma\lambda} \Delta_{\rho}^{(+)} N_{\sigma\lambda}.
\end{equation}
By orthogonality $\det P = 1$ and the only non-zero contributions arise when $\rho$, $\sigma$, $\lambda < 5$ (both $N_{5\lambda} = 0$ and $\Delta_5^{(+)} = 0$).
Thus the only non-zero monopole currents are
\begin{equation}
  M_{\mu5} = \epsilon_{\mu 5\rho\sigma\lambda}\Delta_{\rho}^{(+)} N_{\sigma\lambda},
\end{equation}
which is consistent with the hypercubic expression (\ref{worldline}).

The problem with this approach is that it necessitates gauge fixing (specifically to Lorentz gauge), since one is taking linear combinations of different lattice gauge fields which transform differently under gauge transformations.
For this reason we have not used this $S_5$-symmetric approach in our current measurements of the monopole density.
Instead we have simply used the four $\hatbmu_{\mu}$ (\ref{eq:dirs}) that point along the principal axes of the hypercubic lattice.
We can gain some intuition for the effects of this procedure by considering a similar, simpler system, the two-dimensional XY model.
This system possesses vorticity, defined as the sum around the unit cell of the differences between the phases of spins at adjacent sites.
Our approach is analogous to measuring the total vorticity of two cells, which would average nearby vortex--antivortex pairs to zero, a UV effect.

\begin{figure*}[btp]
  \centering
  \includegraphics[height=\figheight]{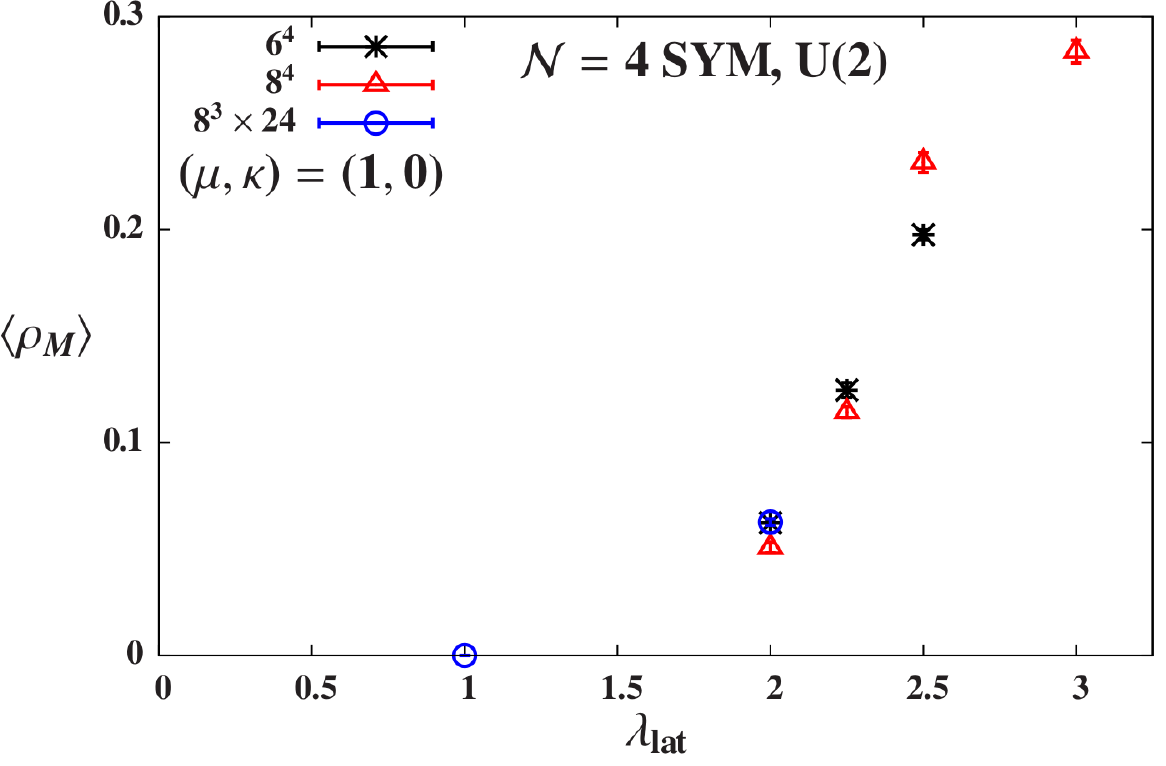}
  \caption{\label{fig:condensation}The density of monopole world lines in the U(1) sector plotted vs.\ the 't~Hooft coupling $\lambda_{\rm lat}$ for fixed $\mu = 1$ indicates that the strong-coupling confinement transition observed in Figs.~\protect\ref{fig:pl_vs_lambda} and \protect\ref{fig:det_vs_lambda} is caused by the compact U(1) degrees of freedom in the U($N$) gauge group.}
\end{figure*}

Results of U(2) simulations are shown in \fig{fig:condensation}, where the mean density of monopole world lines $\rho_M = \frac{1}{4V}\sum_{\vn, \mu}|M_{\mu}(\vn)|$ is plotted as a function of $\lambda_{\rm lat}$ at fixed $\mu = 1$.
The density rises around the same values of $\lambda_{\rm lat}$ where the Polyakov loop and plaquette determinant plunge in Figs.~\ref{fig:pl_vs_lambda} and \ref{fig:det_vs_lambda}.

\subsection{\label{sec:det}Removing the confining phase}
We can suppress the compact U(1) gauge degrees of freedom by adding to the action a term
\begin{equation}
  \label{eq:det_term}
  S_{det} = \kappa \sum_{\cP} |\det \cP - 1|^2,
\end{equation}
where $\kappa$ is a new tunable parameter and $\cP$ is the product of complexified gauge links $\cU_a$ around a fundamental plaquette of the lattice.
This projects the plaquettes from \glN to $\slN$, and SL($N, \Cbb$) does not have a compact U(1) subgroup.
This is not precisely equivalent to reducing the gauge group from U($N$) to SU($N$), but it certainly suppresses the U(1) sector and hence serves the right purpose.

\begin{figure*}[btp]
  \centering
  \includegraphics[height=\figheight]{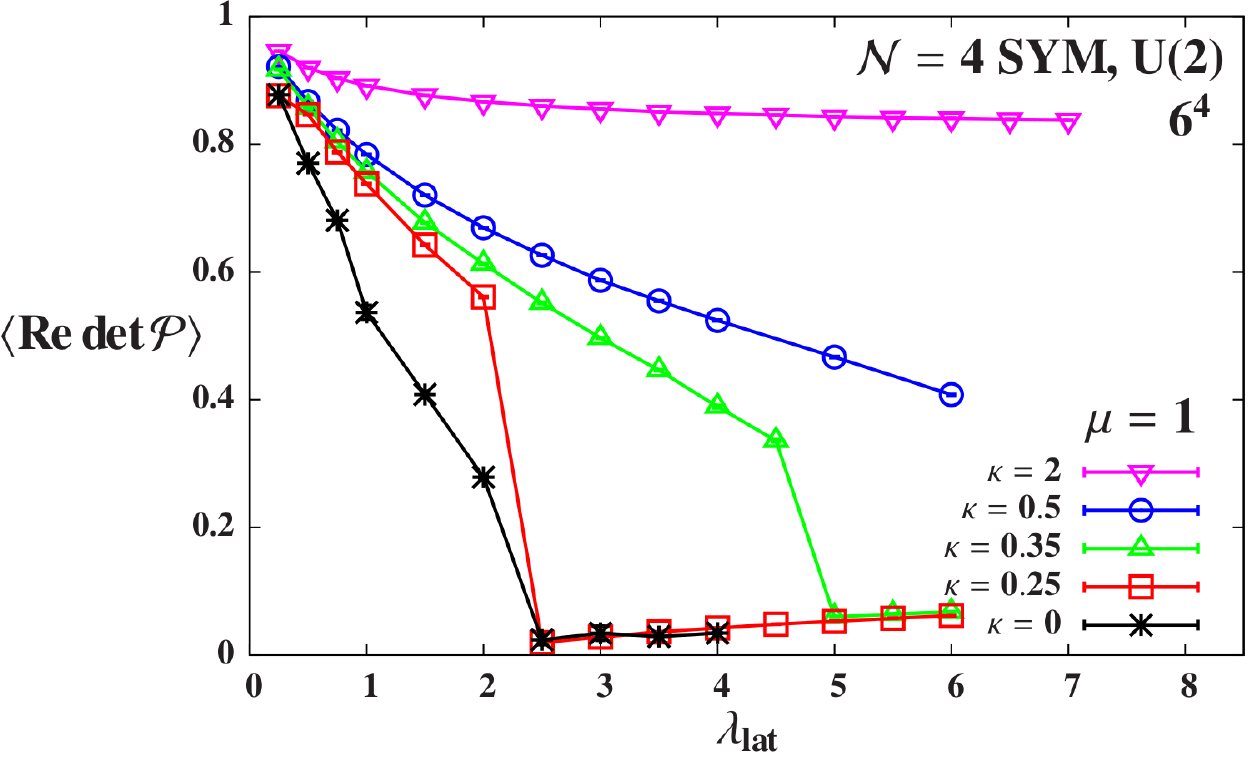}
  \caption{\label{fig:ndet} The real part of the plaquette determinant vs.\ the 't~Hooft coupling $\lambda_{\rm lat}$ on $6^4$ lattices with a variety of $\kappa$ in \protect\eq{eq:det_term} and fixed $\mu = 1$.  As $\kappa$ increases the confinement transition associated with the U(1) sector moves to larger coupling, and disappears entirely for $\kappa \geq 0.5$.  Lines connect points with the same $\kappa$ to guide the eye.}
\end{figure*}

\fig{fig:ndet} shows what happens when we turn on the term (\ref{eq:det_term}) in the action to suppress the U(1) gauge degrees of freedom.
(This figure considers different $\kappa = 0$ ensembles than does \fig{fig:det_vs_lambda}.)
As $\kappa$ increases, the U(1) confinement transition moves to larger $\lambda_{\rm lat}$, disappearing entirely for $\kappa \geq 0.5$.
There is a simple argument explaining this threshold value of $\kappa$: to leading order the determinant term (\ref{eq:det_term}) yields a $2\kappa F_{\mu\nu} F^{\mu\nu}$ lattice action for the U(1) sector, corresponding to compact four-dimensional QED with $\beta_{U(1)} = 2\kappa$.
Thus the known critical value $\beta_{U(1)}^{(c)} = 0.99$ reported in Ref.~\cite{DeGrand:1980eq} implies a critical $\kappa \approx 0.5$ as observed.
Larger values of $\kappa$ correspond to the weak-coupling phase of the U(1) theory.

Like the $\mu$ term in \eq{eq:mass}, the $\kappa$ term (\ref{eq:det_term}) involves only the bosonic fields and thus softly breaks the scalar supersymmetry $\cQ$.
Unlike the relevant coupling $\mu$, $\kappa$ is marginal, as we can see by its appearance in the $F_{\mu\nu} F^{\mu\nu}$ term of the U(1) sector.
In addition, in the continuum limit the U(1) sector in which $\kappa$ breaks supersymmetry decouples from the SU($N$) theory of interest.
This suggests that it may not be necessary to tune $\kappa$ to zero in the continuum limit $1 / L \to 0$, and we intuitively expect its supersymmetry breaking to be mild even without taking the continuum limit.
In Section~\ref{sec:tests} we present some numerical evidence supporting this expectation.

\begin{figure*}[btp]
  \includegraphics[height=\figheight]{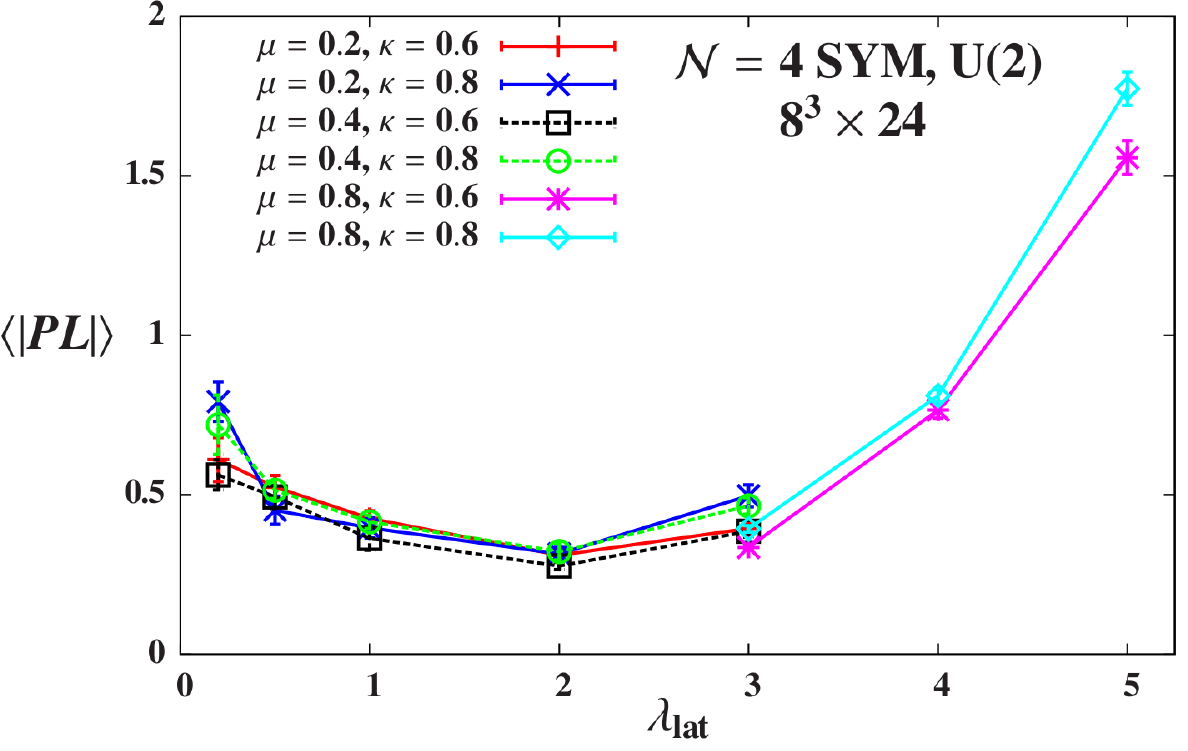}
  \caption{\label{fig:pl824}The absolute value of the Polyakov loop with non-zero $\kappa > 0.5$ on $8^3\times24$ lattices.  The six data sets with $(\mu, \kappa) = (0.2, 0.6)$, (0.2, 0.8), (0.4, 0.6), (0.4, 0.8), (0.8, 0.6) and (0.8, 0.8) are plotted vs.\ the 't~Hooft coupling $\lambda_{\rm lat}$.  Despite the long temporal extent of lattice, $N_t = 24$, the results remain $\cO(1)$ for all investigated couplings, as expected for a deconfined system.  No significant dependence on $\mu$ or $\kappa$ is visible.  At stronger couplings $\lambda_{\rm lat} \gtrsim 3$ larger fluctuations in the non-unitarized link fields cause $\vev{|PL|}$ to increase, and require that we use larger values of $\mu$ to maintain stability. Lines connect points with fixed $(\mu, \kappa)$ to guide the eye.}
\end{figure*}

\begin{figure*}[btp]
  \includegraphics[height=\figheight]{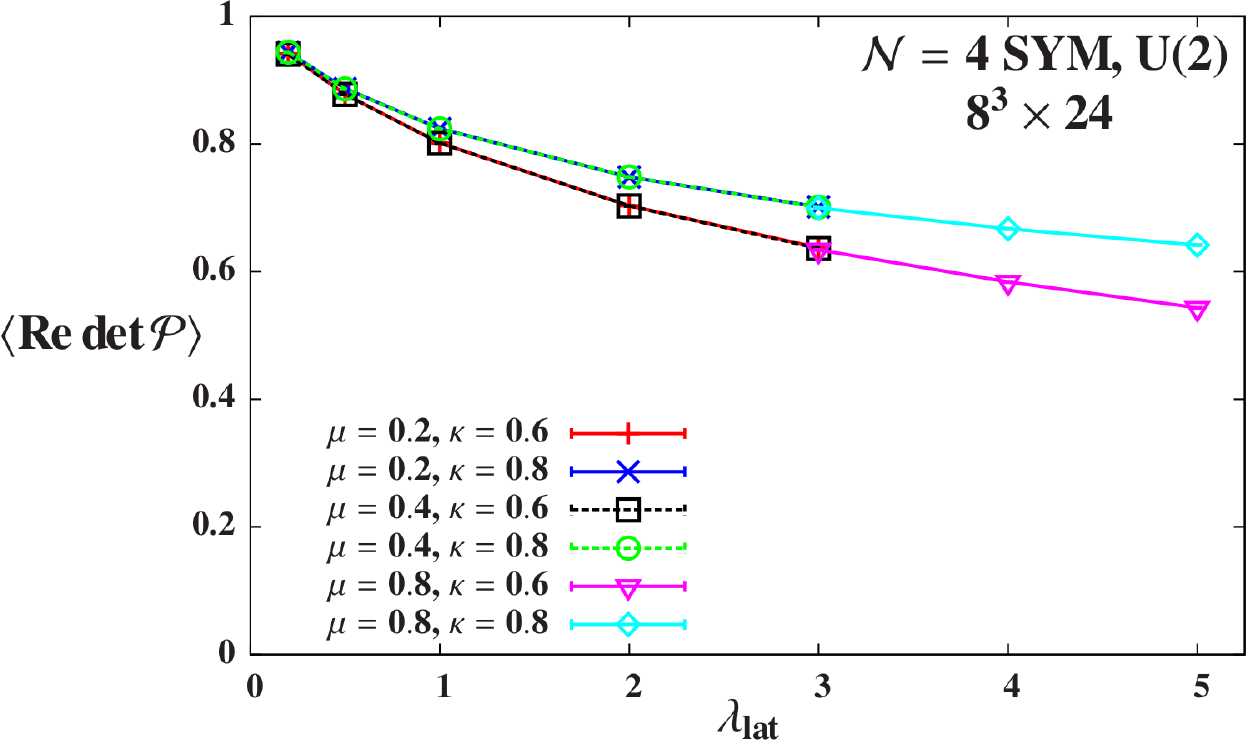}
  \caption{\label{fig:det824}The real part of the plaquette determinant with non-zero $\kappa > 0.5$ on $8^3\times24$ lattices.  The six data sets with $(\mu, \kappa) = (0.2, 0.6)$, (0.2, 0.8), (0.4, 0.6), (0.4, 0.8), (0.8, 0.6) and (0.8, 0.8) are plotted vs.\ the 't~Hooft coupling $\lambda_{\rm lat}$.  The results remain $\cO(1)$ for all investigated couplings, with no visible dependence on $\mu$.  As expected, larger $\kappa$ keeps $\vev{\mbox{Re}\det\cP}$ closer to unity, especially at stronger couplings.  Lines connect points with fixed $(\mu, \kappa)$ to guide the eye.}
\end{figure*}

We conclude this section by presenting the Polyakov loop and plaquette determinant on $8^3\times24$ lattices with non-zero $\kappa > 0.5$, in \fig{fig:pl824} and \fig{fig:det824}, respectively.
These ensembles will be the focus of our following investigations, and they are summarized in Table~\ref{data_sets} in the Appendix.
There is no sign of a transition in these ensembles.
For all investigated couplings $\lambda_{\rm lat} \leq 5$ the Polyakov loop $\vev{|PL|}$ retains the $\cO(1)$ magnitude expected for a deconfined system.
(At stronger couplings $\lambda_{\rm lat} \gtrsim 3$ larger fluctuations in the non-unitarized link fields cause $\vev{|PL|}$ to increase, and require that we use larger values of $\mu$ to maintain stability.)
Note that these lattices possess a much longer temporal extent than those considered in \fig{fig:pl_vs_lambda}.
The real part of the plaquette determinant is now a very smooth function of $\lambda_{\rm lat}$ and remains $\cO(1)$ for all couplings we study.
A plot of the monopole density for these runs would be uninteresting: $\rho_M = 0$ over this range of $\lambda_{\rm lat}$ for these $\kappa = 0.6$ and 0.8.



\section{\label{sec:tests}Numerical evidence for lattice $\cN = 4$ supersymmetry}
In the lattice theory described above, the twisted scalar supersymmetry $\cQ$ is exactly preserved at non-zero lattice spacing only if $\mu = 0$ in \eq{eq:mass} and $\kappa = 0$ in \eq{eq:det_term}.
Stable numerical calculations require both non-zero $\mu$ and $\kappa$, softly breaking the $\cQ$ supersymmetry.
Our choice of anti-periodic temporal boundary conditions for the fermionic (but not bosonic) fields leads to further supersymmetry breaking as a finite-volume effect.
The other 15 supersymmetry charges $\cQ_a$ and $\cQ_{ab}$ are broken by the lattice discretization itself, and must be recovered in the continuum limit.

In this section we present evidence that the breaking of $\cQ$ is under control in our numerical calculations.
We then discuss the fate of the full $\cN = 4$ supersymmetry, developing lattice observables that can be used to monitor the discrete R symmetries mentioned in Section~\ref{sec:continuum}.
We also explore the possibility of a sign problem in our phase-quenched simulations, which omit the phase of the pfaffian from the measure of the path integral.
All indications are that our calculations do not suffer from a sign problem; while not conclusive, this result is quite encouraging.
Finally, in Section~\ref{sec:largeN} we present encouraging initial results from studies of the U($N$) gauge groups with $N = 3$ and 4.
We observe $1 / N^2$ suppression of supersymmetry-breaking effects and no change in the pfaffian phase as $N$ increases.

\subsection{\label{sec:susy}The scalar supersymmetry $\cQ$}
\begin{figure*}[btp]
  \includegraphics[height=\figheight]{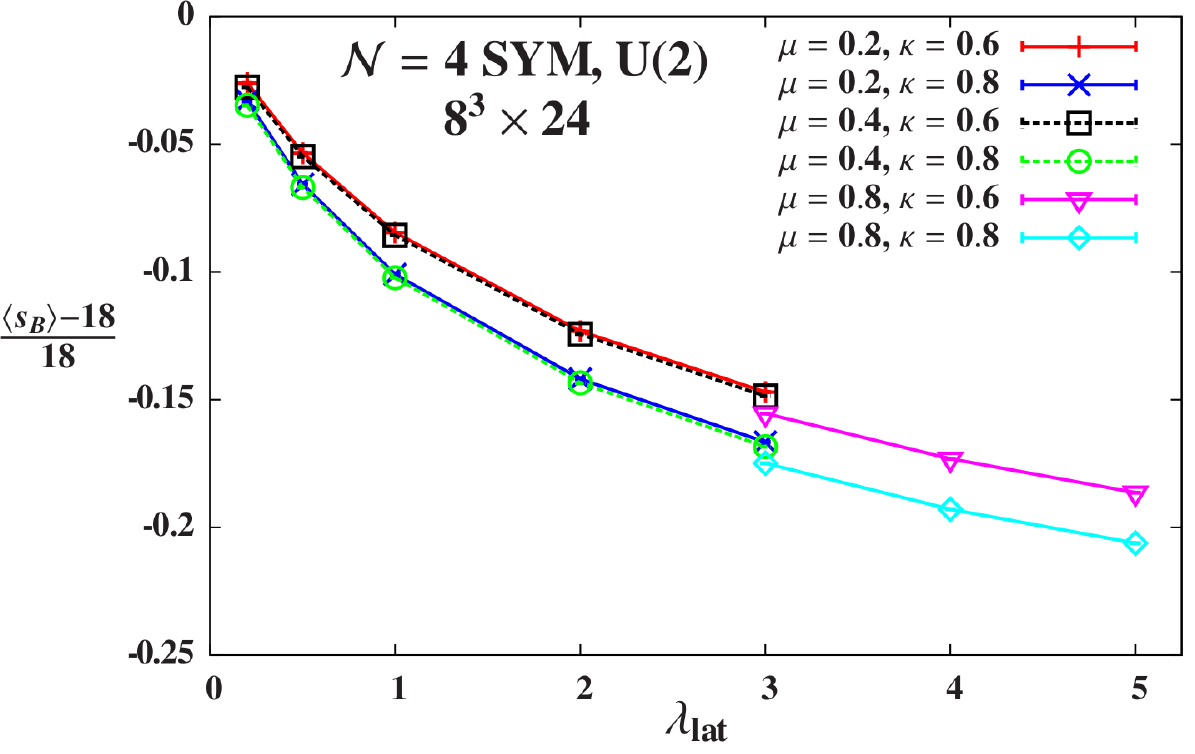}
  \caption{\label{fig:sB} Deviations of the bosonic action $\vev{s_B}$ from its exact supersymmetric value $9N^2 / 2$ serve as a measure of $\cQ$ supersymmetry breaking due to non-zero $\mu$ and $\kappa$.  Six $8^3\times24$ data sets with $(\mu, \kappa) = (0.2, 0.6)$, (0.2, 0.8), (0.4, 0.6), (0.4, 0.8), (0.8, 0.6) and (0.8, 0.8) are plotted vs.\ the 't~Hooft coupling $\lambda_{\rm lat}$.  The results indicate mild $\cO(10\%)$ supersymmetry breaking in these $8^3\times24$ lattice ensembles that are the focus of our current investigations.  As expected, larger $\mu$ and $\kappa$ each increase supersymmetry breaking, though $s_B$ is more sensitive to $\kappa$ than to $\mu$.  The deviations also grow with the coupling, and approach zero in the free-field limit $\lambda_{\rm lat} \to 0$.  Lines connect points with fixed $(\mu, \kappa)$ to guide the eye.}
\end{figure*}

The bosonic action $S_B$ provides an indirect means to explore how badly the scalar supersymmetry $\cQ$ is broken by non-zero $\mu$ and $\kappa$ in our numerical calculations.
Normalizing $S_B$ by the lattice volume, supersymmetry permits exact analytic calculation of $s_B \equiv S_B / V = 9N^2 / 2$ for gauge group U($N$).
In \fig{fig:sB} we plot $(\vev{s_B} - 18) / 18$ for $N = 2$ on $8^3\times 24$ lattices, as a function of the 't~Hooft coupling $\lambda_{\rm lat}$ for several values of $(\mu, \kappa)$.
As expected, deviations from the exact supersymmetric result increase with each of $\mu$, $\kappa$ and $\lambda_{\rm lat}$, reaching $\sim$20\% for the strongest coupling we consider, $\lambda_{\rm lat} = 5$.
For $\lambda_{\rm lat} \leq 1$ the deviations are $\sim$10\% at most.
We see that $s_B$ is much more sensitive to $\kappa$ than to $\mu$: the effect of doubling $\mu = 0.2 \to 0.4$ is negligible compared to that of increasing $\kappa$ from 0.6 to 0.8.
This is consistent with Figure~4 in Ref.~\cite{Catterall:2012yq}, which shows $\vev{s_B} / 18 \gtrsim 0.98$ for all $\lambda_{\rm lat} \leq 2.6$ and $\mu \leq 1$ when the $\kappa$ coupling (\ref{eq:det_term}) is not included in the action.

An alternate measure of $\cQ$-breaking is the vacuum expectation value of the supersymmetry transformation of a suitable local operator, $\vev{\cQ \cO}$.
As discussed in Ref.~\cite{Catterall:2013roa}, the exact $\cQ$ supersymmetry of our lattice action $S_0$ (\ref{eq:lat_act}) without the terms (\ref{eq:mass}) and (\ref{eq:det_term}) implies the Ward identity $\vev{\cQ \cO} = 0$.
Because non-zero $\mu$ and $\kappa$ softly break the supersymmetry $\cQ$, in numerical computations we will find $\vev{\cQ \cO}$ to be non-zero.
These Ward identity violations indicate how badly supersymmetry is broken.

In order to measure $\vev{\cQ \cO}$ on the lattice, we need to choose an appropriate operator $\cO$, which must be fermionic so that its $\cQ$ transformation is bosonic.
One simple operator that does not already appear in our action takes the continuum form
\begin{equation}
  \cO = \mbox{Tr}\left[\eta \sum_a \cA_a \cAb_a\right].
\end{equation}
Acting upon $\cO$ with the supersymmetry $\cQ$ produces
\begin{equation}
  \cQ \cO = \mbox{Tr}\left[\sum_b \left[\cDb_b, \cD_b\right] \sum_a \cA_a \cAb_a\right] - \mbox{Tr}\left[\eta \sum_a \psi_a \cAb_a\right],
\end{equation}
where we have used the equation of motion for the auxiliary field $\cQ\, \eta = d = \sum_b \left[\cDb_b, \cD_b\right]$.
The first term in the right-hand side above is constructed entirely from the complexified gauge fields, while the second involves the $\eta\psi_a$ fermion bilinear.
The relative negative sign comes from anti-commuting $\cQ$ through $\eta$.

\begin{figure*}[btp]
  \includegraphics[height=\figheight]{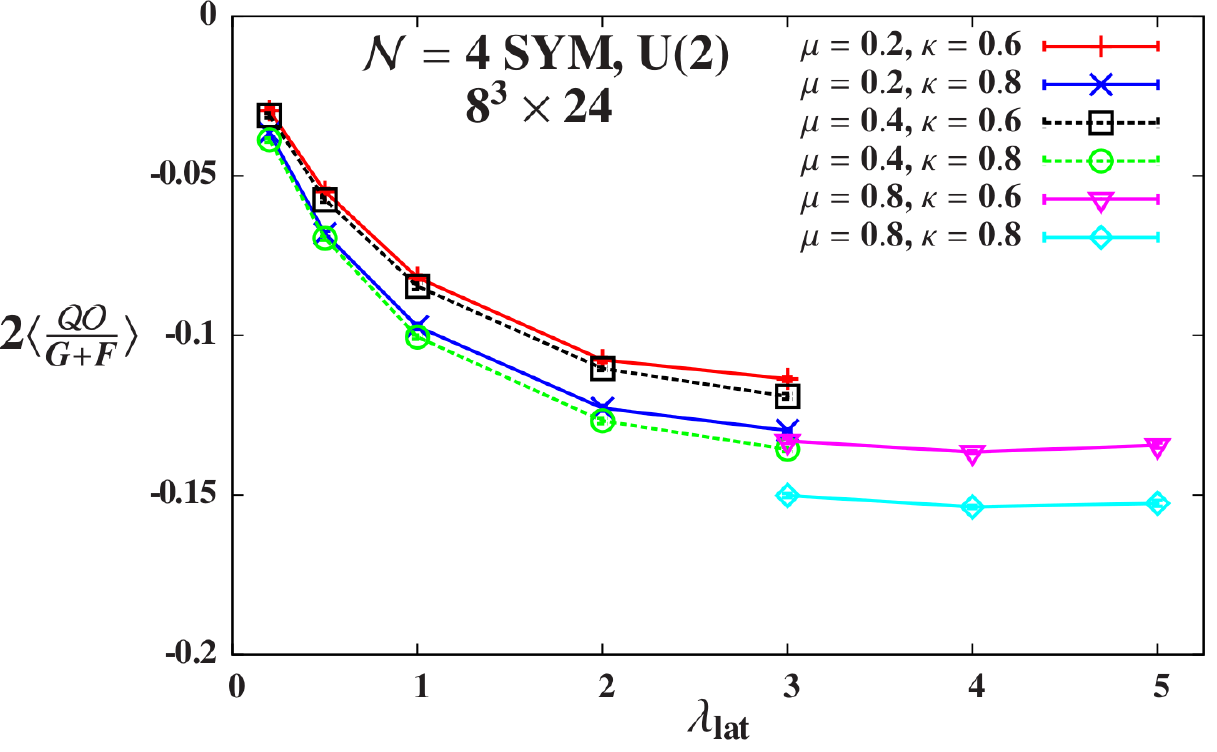}
  \caption{\label{fig:susy_breaking} Violations of the Ward identity $\vev{\cQ \cO} = 0$ as defined by \protect\eq{eq:susy_rel} serve as a measure of $\cQ$ supersymmetry breaking due to non-zero $\mu$ and $\kappa$.  Six $8^3\times24$ data sets with $(\mu, \kappa) = (0.2, 0.6)$, (0.2, 0.8), (0.4, 0.6), (0.4, 0.8), (0.8, 0.6) and (0.8, 0.8) are plotted vs.\ the 't~Hooft coupling $\lambda_{\rm lat}$.  The results indicate mild $\cO(10\%)$ supersymmetry breaking in these $8^3\times24$ lattice ensembles that are the focus of our current investigations.  As expected, larger $\mu$ and $\kappa$ each increase supersymmetry breaking, though $\vev{\cQ \cO}$ is more sensitive to $\kappa$ than to $\mu$.  The Ward identity violations also grow with the coupling, and approach zero in the free-field limit $\lambda_{\rm lat} \to 0$.  Lines connect points with fixed $(\mu, \kappa)$ to guide the eye.}
\end{figure*}

Transcribed onto the lattice, $\left[\cDb_b, \cD_b\right] \to \left[\cU_b(\vn) \cUb_b(\vn) - \cUb_b\left(\vn - \hatbmu_b\right) \cU_b\left(\vn - \hatbmu_b\right)\right]$ and
\begin{equation}
  \label{eq:susy_latop}
  \cQ \cO = \mbox{Tr}\left[\sum_b \left[\cU_b\cUb_b  - \cUb_b\cU_b\right] \sum_a \cU_a \cUb_a\right] - \mbox{Tr}\left[\eta \sum_a \psi_a \cUb_a\right].
\end{equation}
Averaging over the lattice volume, the vacuum expectation value of each term in the right-hand side of \eq{eq:susy_latop} grows rapidly as $\lambda_{\rm lat}$ increases.
We normalize their difference by the average of the two terms, plotting in \fig{fig:susy_breaking} the quantity
\begin{equation}
  \label{eq:susy_rel}
  \frac{\mbox{Tr}\left[\sum_b \left[\cU_b\cUb_b  - \cUb_b\cU_b\right] \sum_a \cU_a \cUb_a\right] - \mbox{Tr}\left[\eta \sum_a \psi_a \cUb_a\right]}{\frac{1}{2}\left\{\mbox{Tr}\left[\sum_b \left[\cU_b\cUb_b  - \cUb_b\cU_b\right] \sum_a \cU_a \cUb_a\right] + \mbox{Tr}\left[\eta \sum_a \psi_a \cUb_a\right]\right\}} \equiv 2\frac{G - F}{G + F} = 2\frac{\cQ \cO}{G + F},
\end{equation}
using the shorthand ``$G$'' and ``$F$'' to refer to the gauge and fermion-bilinear terms, respectively.
We find that the violations of this Ward identity are quite similar to the deviations of the bosonic action from its exact value, though slightly less sensitive to $\lambda_{\rm lat}$ and slightly more sensitive to $\mu$.
By considering 2--3 values of $\mu$, each with two $\kappa > 0.5$, we can observe that supersymmetry breaking appropriately decreases as $\mu$ and $\kappa$ get smaller.
We have not yet attempted to extrapolate these results to the supersymmetric continuum limit $(1 / L, \mu, \kappa) \to (0, 0, 0)$.

In both observables, additional supersymmetry breaking from finite-volume effects is negligible compared to that due to non-zero $\mu$ and (especially) $\kappa$.
Considering fixed $(\lambda_{\rm lat}, \mu, \kappa) = (1, 1, 1)$, Table~\ref{susy_vol_scaling} shows that the bosonic action and Ward identity violations barely vary as the total lattice volume increases by an order of magnitude from $8^3\times24$ to $16^3\times32$.
Even a very small $4^4$ lattice volume produces only percent-level corrections to the dominant contributions from the large $\kappa = 1$ and $\mu = 1$.
The results are also insensitive to the temporal extent of the lattice, despite the different temporal boundary conditions we impose for the bosonic and fermionic fields.

\begin{table}[tbp]
  \caption{\label{susy_vol_scaling} Deviations from the exact supersymmetric bosonic action $\vev{s_B}$, and violations of the Ward identity $\vev{\cQ \cO} = 0$ as defined by \protect\eq{eq:susy_rel}, for different lattice volumes at fixed $(\lambda_{\rm lat}, \mu, \kappa) = (1, 1, 1)$ and gauge group U(2).  The results indicate that additional supersymmetry breaking from finite-volume effects is negligible compared to that due to non-zero $\mu$ and $\kappa$.  Even a very small $4^4$ lattice volume produces only percent-level corrections compared to results from $8^3\times24$, $12^3\times24$ and $16^3\times32$ volumes.}
  \centering
  \renewcommand\arraystretch{1.2}  
  \begin{tabular}{|c|c|c|}
    \hline
    Volume            & $\frac{\vev{s_B} - 18}{18}$ & $2\vev{\frac{\cQ \cO}{G + F}}$  \\\hline
    ~$16^3\times32$~  & ~$-$0.12254(4)~             & ~$-$0.12785(23)~                \\
    $12^3\times24$    &  $-$0.12246(12)             &  $-$0.12721(60)                 \\
    $8^3\times24$     &  $-$0.12269(16)             &  $-$0.12701(61)                 \\\hline
    $4^4$             &  $-$0.1242(6)               &  $-$0.1300(26)                  \\
    \hline
  \end{tabular}
\end{table}

In addition to monitoring the effects of non-zero $\mu$ and $\kappa$ in the $8^3\times24$ lattice ensembles that are the focus of our current investigations, we can also confirm that $\cQ$ is indeed restored in the limit $(\mu, \kappa) \to (0, 0)$.
This was done in Ref.~\cite{Catterall:2012yq} for the case $\kappa = 0$: Figure~4 in that reference shows that the normalized bosonic action $\vev{s_B} / 18$ approaches unity as the bosonic mass decreases, indicating recovery of exact $\cQ$ supersymmetry.
Here we extend this investigation to non-zero $\kappa$, scanning a broad range of $0.2 \leq \mu \leq 1$ and $0.1 \leq \kappa \leq 1$ on $4^4$ lattices with fixed $\lambda_{\rm lat} = 1$.

\begin{figure*}[btp]
  \includegraphics[height=\figheight]{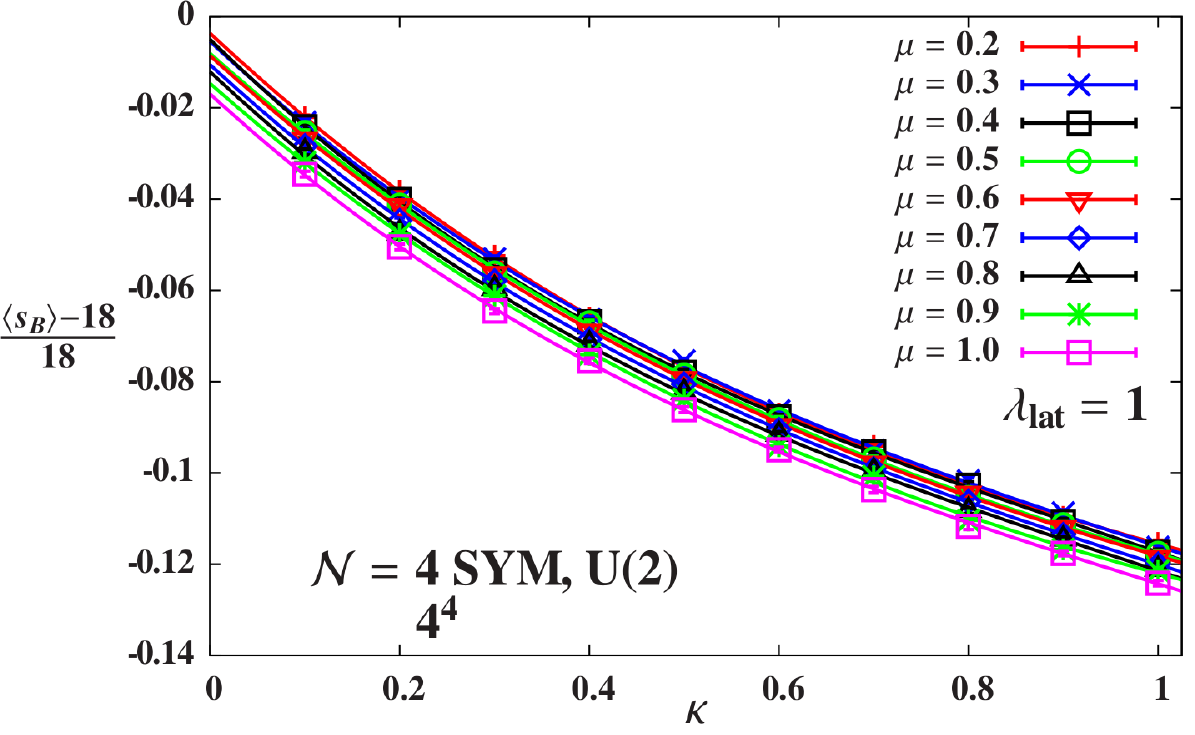}
  \caption{\label{fig:sB_extrap} Deviations of the bosonic action $\vev{s_B}$ from its exact supersymmetric value $9N^2 / 2$, as in \protect\fig{fig:sB}.  Nine $4^4$ data sets with $0.2 \leq \mu \leq 1$ are plotted vs.\ $\kappa$ for fixed $\lambda_{\rm lat} = 1$.  While the results again show little dependence on $\mu$, the deviations steadily grow as $\mu$ increases, as expected.  The lines are cubic $\kappa \to 0$ extrapolations.  A subsequent quadratic $\mu \to 0$ extrapolation of their intercepts produces $\lim_{(\mu, \kappa) \to (0, 0)}\vev{s_B} = 17.956(37)$, consistent with the restoration of the $\cQ$ supersymmetry up to minor effects attributable to the small lattice volume.}
\end{figure*}

\fig{fig:sB_extrap} shows the resulting deviations of the bosonic action $\vev{s_B}$ from its exact supersymmetric value $9N^2 / 2$.
With $\mu$ fixed, cubic $\kappa \to 0$ extrapolations fit the data well, with $\chi^2 / \mbox{d.o.f.} \lesssim 1$.
A subsequent quadratic $\mu \to 0$ extrapolation produces $\lim_{(\mu, \kappa) \to (0, 0)}\vev{s_B} = 17.956(37)$.
If instead we consider the points with $\mu = \kappa$ and carry out a single cubic extrapolation, we find $\lim_{(\mu, \kappa) \to (0, 0)}\vev{s_B} = 17.953(44)$.
Both of these values are consistent with the restoration of the $\cQ$ supersymmetry in the $(\mu, \kappa) \to (0, 0)$ limit.
According to Table~\ref{susy_vol_scaling}, the remaining differences between $\vev{s_B}$ and its exact supersymmetric value are the size we should expect to result from working on such a small $4^4$ lattice.

\begin{figure*}[btp]
  \includegraphics[height=\figheight]{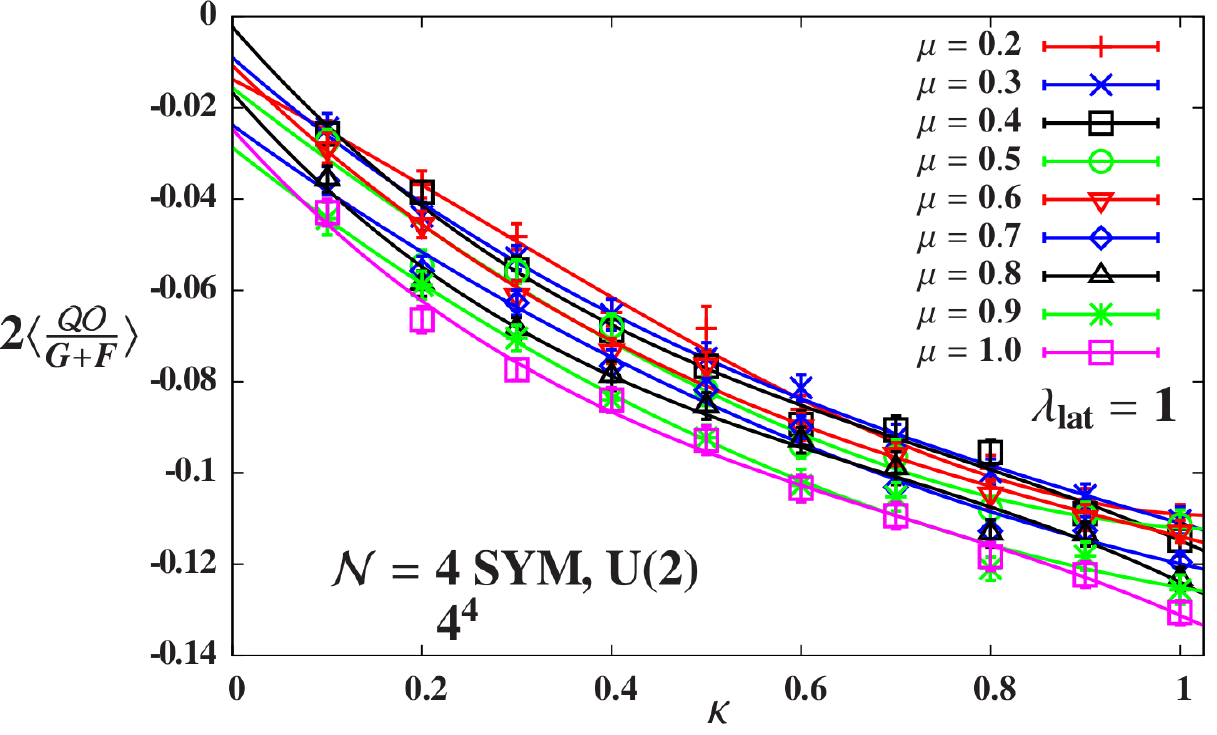}
  \caption{\label{fig:susy_extrap} Violations of the Ward identity $\vev{\cQ \cO} = 0$ as defined by \protect\eq{eq:susy_rel}, as in \protect\fig{fig:susy_breaking}.  Nine $4^4$ data sets with $0.2 \leq \mu \leq 1$ are plotted vs.\ $\kappa$ for fixed $\lambda_{\rm lat} = 1$.  Although the results are somewhat noisy, the Ward identity violations grow as $\mu$ increases, as expected.  The lines are cubic $\kappa \to 0$ extrapolations.  A subsequent quadratic $\mu \to 0$ extrapolation of their intercepts produces $\lim_{(\mu, \kappa) \to (0, 0)}\vev{\cQ \cO} = -0.011(10)$, consistent with the restoration of the $\cQ$ supersymmetry up to minor effects attributable to the small lattice volume.}
\end{figure*}

Similarly, \fig{fig:susy_extrap} plots violations of the Ward identity $\vev{\cQ \cO} = 0$ as defined by \protect\eq{eq:susy_rel}, for the same $4^4$ lattice ensembles with $0.2 \leq \mu \leq 1$, $0.1 \leq \kappa \leq 1$ and fixed $\lambda_{\rm lat} = 1$.
These results are noisier than those for the bosonic action, but lead to the same conclusion.
Cubic $\kappa \to 0$ extrapolations with fixed $\mu$, followed by a quadratic $\mu \to 0$ extrapolation, produce $\lim_{(\mu, \kappa) \to (0, 0)}\vev{\cQ \cO} = -0.011(10)$, while a single cubic extrapolation with $\mu = \kappa$ leads to $\lim_{(\mu, \kappa) \to (0, 0)}\vev{\cQ \cO} = 0.006(13)$.
Again, Table~\ref{susy_vol_scaling} indicates that the remaining Ward identity violations in the $(\mu, \kappa) \to (0, 0)$ limit may be attributed to the small $4^4$ lattice volume.

The insensitivity to $1 / L$ indicated by Table~\ref{susy_vol_scaling} is likely a simple reflection of the fact that both observables are expectation values of local operators averaged over the lattice volume.
In order to explore $\cQ$ supersymmetry breaking in the continuum limit, we may need to search for appropriate extended operators $\vev{\cQ \cO(x - y)}$, which would allow us to probe a range of distance scales $r = |x - y|$.
In the next subsection we will see that observables sensitive to the restoration of the other 15 supersymmetries $\cQ_a$ and $\cQ_{ab}$ naturally have such an extended character.

\subsection{\label{sec:other_susies}The other 15 supersymmetries $\cQ_a$ and $\cQ_{ab}$}
The fate of the 15 broken supersymmetry generators on the lattice is obviously of crucial importance.
As discussed in Section~\ref{sec:continuum}, we must recover these supersymmetries in the continuum limit in order to claim that we are faithfully simulating $\cN = 4$ SYM.
Recently Ref.~\cite{Catterall:2013roa} showed how restoration of the full $\cN = 4$ supersymmetry follows from preservation of $\cQ$ and a set of discrete R symmetries, subgroups of the continuum SU(4)$_R$ symmetry.
These discrete R symmetries may be labeled in a way analogous to the twisted supersymmetries, as the set $\{R_a, R_{ab}\}$.
As an example, the symmetry transformation $R_a$ acts on the fields according to
\begin{align}
  & R_a \eta = 2 \psi_a      & & R_a \psi_a = \frac{1}{2} \eta                          & & R_a \psi_b = -\chi_{ab} & &                     \cr
  & R_a \chi_{ab} = - \psi_b & & R_a \chi_{bc} = \frac{1}{2} \epsilon_{bcagh} \chi_{gh} & &                         & & \label{eq:R_a}      \\
  & R_a \cD_a = \cD_a        & & R_a \cDb_a = \cDb_a                                    & & R_a \cD_b = \cDb_b      & & R_a \cDb_b = \cD_b, \nn
\end{align}
where ``$a$'' is a fixed index and $b, c, g, h \ne a$ represent the other indices.
This symmetry of the continuum twisted action can be used to derive the $\cQ_a$ transformations of the fields; morally speaking, $\cQ_a = R_a \cQ$.
Similarly there are ten other discrete R symmetries $R_{ab}$ that yield $\cQ_{ab}$.
The main point, described in detail in Ref.~\cite{Catterall:2013roa}, is that recovery of any of these R symmetries enforces the correct coefficients on the long distance effective action (\ref{eq:S_eff}), and thus the recovery of the full $\cN = 4$ supersymmetry.
Therefore we can test for the restoration of all $\cQ_a$ and $\cQ_{ab}$ simply by performing a single $R_a$ transformation on a gauge invariant observable.

From \eq{eq:R_a}, we see that the $R_a$ transformation of the continuum complexified gauge fields is
\begin{align}
  & R_a \cA_a = \cA_a   &
  & R_a \cAb_a = \cAb_a &
  & R_a \cA_b = \cAb_b  &
  & R_a \cAb_b = \cA_b
\end{align}
for all $b \ne a$.
Recalling that the lattice link fields $\cU_a = \Ibb + \cA_a + \cO(a)$ in the continuum limit, we define a lattice $R_a$ transformation as
\begin{align}
  \label{Ratrexp}
  & R_a \cU_a = \cU_a         &
  & R_a \cUb_a = \cUb_a       &
  & R_a \cU_b = \cUb_b^{-1}   &
  & R_a \cUb_b = \cU_b^{-1}.
\end{align}
This definition ensures that the lattice $R_a$ transformation commutes with lattice gauge invariance.
Thus a simple test of the $R_a$ invariance is to consider the plaquette
\begin{align}
  \cP_{ab} & = \tr\left[\cU_a(x) \cU_b(x + \hatbe_a) \cUb_a(x + \hatbe_b) \cUb_b(x)\right]  \nn \\
  \implies R_a \cP_{ab} & = \tr\left[\cU_a(x) \cUb_b^{-1}(x + \hatbe_a) \cUb_a(x + \hatbe_b) \cU_b^{-1}(x)\right] \equiv \cPtwiddle_{ab} \label{eq:Rplaq_rel}.
\end{align}
Because our gauge links are non-unitary elements of $\glN$, $\cPtwiddle_{ab} \ne \cP_{ab}$.
However, $\cPtwiddle_{ab}$ still involves closed paths on the lattice and hence is gauge invariant.

\begin{figure*}[btp]
  \centering
  \includegraphics[height=\figheight]{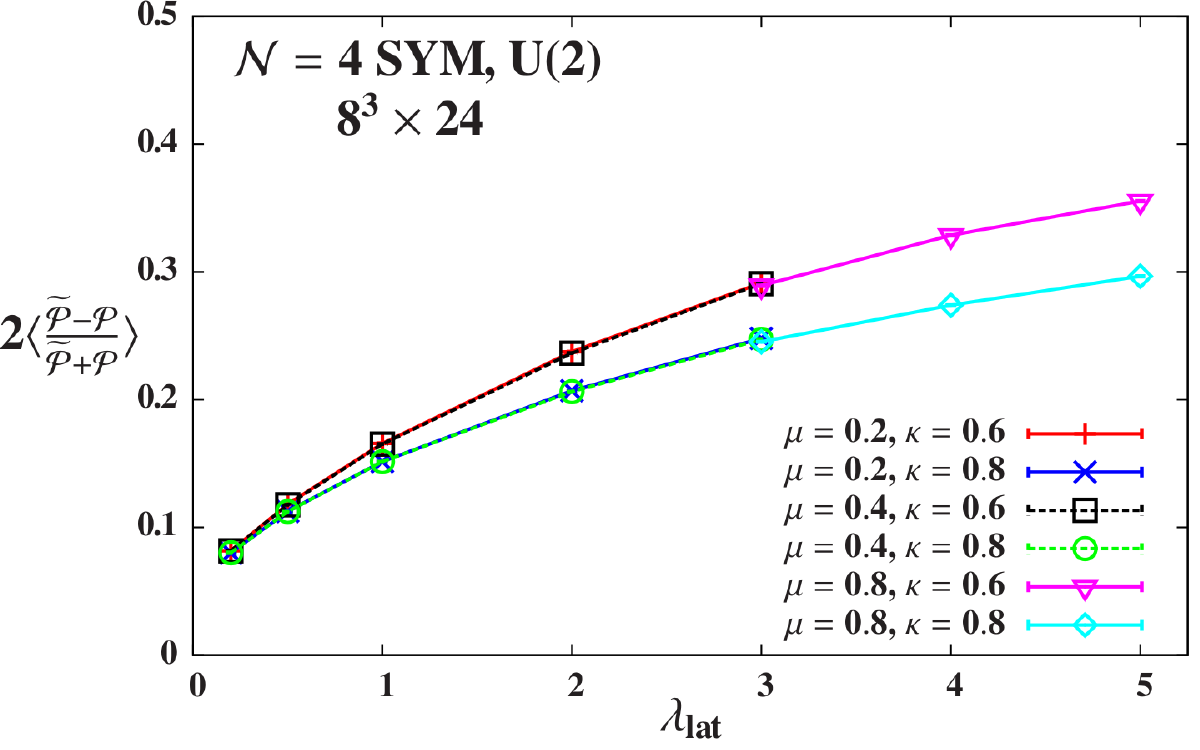}
  \caption{\label{fig:Rplaq} Violations of $R_a$ symmetry in the plaquette, based on \protect\eq{eq:Rplaq_rel}.  Six $8^3\times24$ data sets with $(\mu, \kappa) = (0.2, 0.6)$, (0.2, 0.8), (0.4, 0.6), (0.4, 0.8), (0.8, 0.6) and (0.8, 0.8) are plotted vs.\ the 't~Hooft coupling $\lambda_{\rm lat}$.  The results indicate mild $\cO(10\%)$ R symmetry breaking in these $8^3\times24$ lattice ensembles that are the focus of our current investigations.  No dependence on $\mu$ is visible, while larger $\kappa$ reduces R symmetry breaking, which may hint at a potential connection to the U(1) sector.  The $R_a$ violations also grow with the coupling, and approach zero in the free-field limit $\lambda_{\rm lat} \to 0$.  Lines connect points with fixed $(\mu, \kappa)$ to guide the eye.}
\end{figure*}

In \fig{fig:Rplaq} we show the difference between $\cP_{ab}$ and $\cPtwiddle_{ab}$, normalized by their average value.
At non-zero $\lambda_{\rm lat}$, the $R_a$ symmetry is violated at $\cO(10\%)$ relative to the size of the plaquette.
This level of violation is smaller than one might anticipate, not much worse than we observed for the scalar $\cQ$ supersymmetry in \fig{fig:susy_breaking}.
After all, we only expect the continuum symmetries to be restored at long distances, while the 1$\times$1 Wilson loop is certainly not a long distance quantity, and is heavily influenced by lattice artifacts.
We are currently examining the transformation for larger Wilson loops.
We have also not yet attempted to fine-tune the two independent coefficients of the effective action (\ref{eq:S_eff}), all of which are marginal with the exception of the ``$\beta$'' term (which would be added as a counterterm and may be forbidden by topological arguments~\cite{Catterall:2014mha}).

One novel feature of \fig{fig:Rplaq} is that (unlike the $\cQ$ supersymmetry breaking discussed in the previous subsection) the R symmetry breaking decreases as $\kappa$ increases, which may hint at a potential connection to the U(1) sector.
Recall, however, that restoration of $\cQ_a$ and $\cQ_{ab}$ relies on both the $\cQ$ supersymmetry as well as the $R_a$ symmetry.
Consequently, and as should be expected, the 15 supersymmetries $\cQ_a$ and $\cQ_{ab}$ can not be restored simply by increasing $\kappa$.

While these results are encouraging evidence that our lattice system simulates $\cN = 4$ SYM to a good approximation, more work is obviously required to directly confirm that we recover all necessary symmetries in the $1 / L \to 0$ continuum limit.
In addition to considering larger Wilson loops, we must also study larger lattice volumes.
The results of these investigations will be reported in a future publication.

\subsection{The phase of the pfaffian}
After the redistribution of the Majorana fermion fields according to the prescription of topological twisting, the integrations over these real fermionic components produce a pfaffian, which for any given gauge field configuration is not manifestly real.
We omit the pfaffian phase from the integration measure, including only the absolute value of the pfaffian.
Such ``phase quenching'' leads to a drastic computational simplification, as it allows us to perform simulations by means of pseudofermions $\Phi$, based on the action $\Phi^{\dag} \left(D^{\dag} D\right)^{-1 / 4} \Phi$, where $D$ is the matrix whose pfaffian defines the true measure~\cite{Catterall:2011cea}.

In principle, the true expectation values $\vev{\cO}_{rw}$ can be reconstructed from phase-quenched measurements via reweighting,
\begin{equation}
  \vev{\cO}_{rw} = \frac{\vev{\cO e^{i\alpha}}}{\vev{e^{i\alpha}}}
\end{equation}
\begin{align}
  \vev{\cO} & = \frac{\int[d\cU] \cO e^{-S_B}\ |\pf D|}{\int[d\cU] e^{-S_B}\ |\pf D|} &
  \vev{\cO}_{rw} = \frac{\int[d\cU] \cO e^{-S_B}\ \pf D}{\int[d\cU] e^{-S_B}\ \pf D},
\end{align}
where $\pf D = |\pf D| e^{i\alpha}$.
If the pfaffian is real and positive at non-zero lattice spacing, then $\vev{e^{i\alpha}} = 1$ and phase reweighting has no effect, $\vev{\cO}_{rw} = \vev{\cO}$.
On the other hand, if $\alpha$ fluctuates far from zero on a significant fraction of the gauge configurations generated through phase-quenched importance sampling, then $\vev{e^{i\alpha}}$ may be consistent with zero and reweighting itself breaks down.
Such a sign problem would place a major obstacle in the way of numerical simulation of the theory.

There are some reasons to suspect that the sign is probably not a problem.
First, the pfaffian of the 16-supercharge model dimensionally reduced to zero dimensions can be proven to be real and positive definite for the SU(2) gauge group~\cite{Krauth:1998xh}.
Next, consider the 16-supercharge model in two space-time dimensions.
Even though this system admits a complex pfaffian as in four dimensions, extensive numerical evidence indicates that there is no sign problem even at non-zero lattice spacing (and certainly in the continuum limit)~\cite{Hanada:2010qg, Catterall:2011aa, Mehta:2011ud, Galvez:2012sv}.
Finally, the results in Section~\ref{sec:susy} provide some indirect numerical evidence for the absence of a sign problem in our lattice formulation of four-dimensional $\cN = 4$ SYM.
Specifically, we observed that in the limit $(\mu, \kappa) \to (0, 0)$ the bosonic action approaches its exact supersymmetric value, and violations of the Ward identity $\vev{\cQ \cO} = 0$ as defined by \eq{eq:susy_rel} vanish.
While we only presented results for a single value of the 't~Hooft coupling $\lambda_{\rm lat} = 1$, our previous study~\cite{Catterall:2012yq} reported similar findings for fixed $\kappa = 0$ across a wide range of $\lambda_{\rm lat}$.
One would not have expected such behavior from a theory with a sign problem.

Here we investigate the sign issue in more detail, initiating a study of the pfaffian phase as a function of the lattice volume.
To carry out this work, we have developed new parallel software implementing the algorithm in Ref.~\cite{Catterall:2003ae}, which will be presented in a future publication~\cite{Schaich:2014}.\footnote{The software can currently be obtained through usqcd.org.}
Despite computational advances, we are still limited to small lattice volumes.
Direct measurement of the pfaffian phase is an extremely demanding computation, far more expensive than generating phase-quenched gauge configurations through the rational hybrid Monte Carlo algorithm.
Its cost scales $\sim\!N_{\Psi}^3$, where $N_{\Psi}$ is the number of elements in the fermionic fields.
As an illustration, the largest system on which we measure the pfaffian, a $4^3\times 6$ lattice with gauge group U(2), has $N_{\Psi} = 24,576$.
Our algorithm requires $\left(N_{\Psi} / 2\right)^2 \approx 151\times 10^6$ applications of the fermion operator to compute the pfaffian for a single $4^3\times 6$ gauge configuration, and uses almost 10,000~MB of memory.
Running on 16 cores to minimize the time to solution, each computation lasts for approximately eight days (not including overhead for optional checkpointing).

\begin{figure*}[btp]
  \centering
  \includegraphics[height=\figheight]{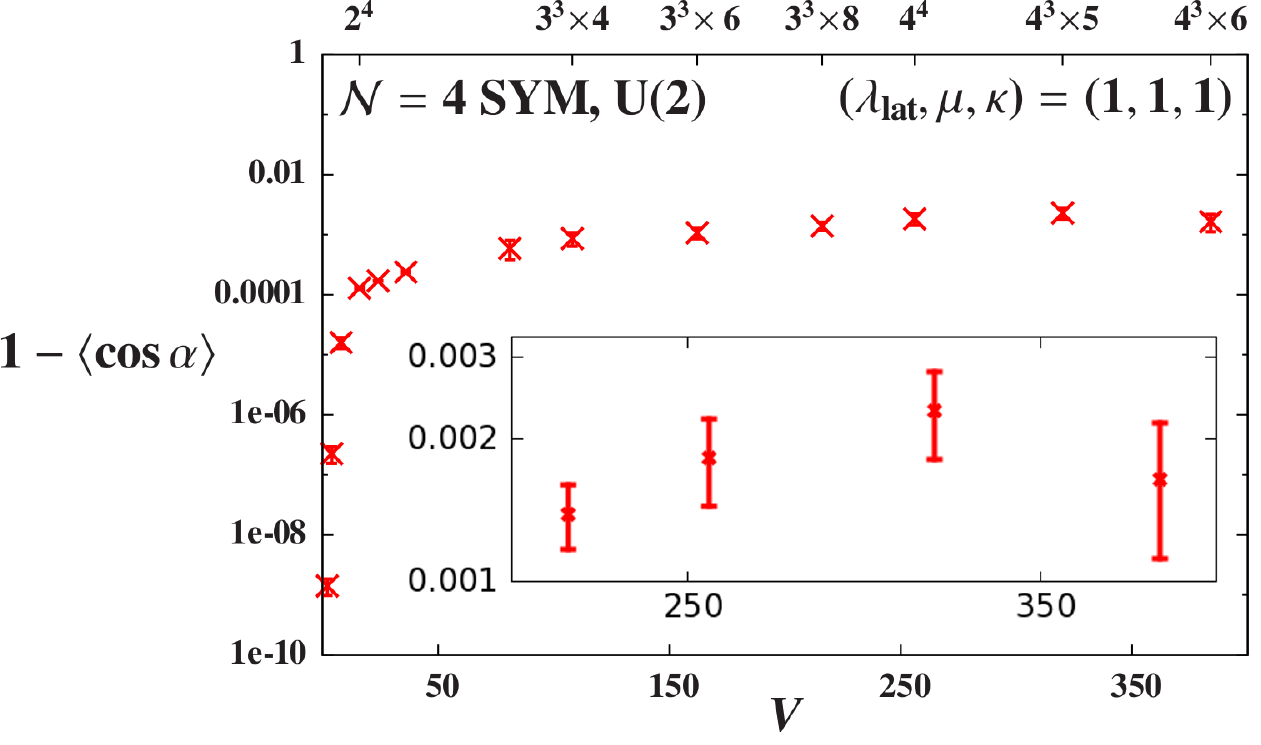}
  \caption{\label{fig:pfaffian} Semi-log plot of $1 - \vev{\cos\alpha}$ vs.\ lattice volume, where $\alpha$ is the phase of the pfaffian.  All points are for the U(2) gauge group with fixed $(\lambda_{\rm lat}, \mu, \kappa) = (1, 1, 1)$.  The inset zooms in on the four largest-volume results, for $3^3\times8$, $4^4$, $4^3\times5$ and $4^3\times6$.  We find that the phase of the phaffian is small, $1 - \vev{\cos\alpha} \lesssim 0.003$, and does not grow on the larger volumes with $L > 3$.}
\end{figure*}

Our results for the pfaffian phase $\alpha$ are shown in Fig.~\ref{fig:pfaffian}, considering fixed $(\lambda_{\rm lat}, \mu, \kappa) = (1, 1, 1)$.
We find that $\vev{e^{i\alpha}} \approx 1$ for all volumes we can investigate, and therefore plot $1 - \vev{\re \left(e^{i\alpha}\right)} = 1 - \vev{\cos\alpha}$ vs.\ the lattice volume on a semi-log scale.
When $L = 1$ in any direction, we effectively consider dimensionally-reduced theories for which $1 - \vev{\cos\alpha}$ is dramatically suppressed.
Although the phase is larger for truly four-dimensional systems, $1 - \vev{\cos\alpha}$ is still small, at most 0.003.
More significantly, the phase angle does not grow noticeably on the largest volumes for which we are able to measure the pfaffian: all of our results with $L > 3$ are constant within uncertainties.
While these initial results on small volumes do not guarantee that our $8^3\times24$ systems are safe from the potential sign problem, they are certainly encouraging, especially given the absence of a similar sign problem in two dimensions.
Further investigations will be presented in a future publication focused on the potential sign problem in our system.

\subsection{\label{sec:largeN}A first look at the large-$N$ limit}
All numerical results discussed above consider the gauge group U(2).
Continuum studies of $\cN = 4$ SYM, in contrast, tend to be anchored in the large-$N$ limit.
Moving to U($N$) with $N > 2$ will be central to our future lattice investigations.
In this subsection we take a first look at larger $N$, specifically the U(3) and U(4) gauge groups.
So far we have analyzed only small lattice volumes for these systems, no larger than $4^4$.
While we have started generating $8^3\times24$ lattice ensembles for $N = 3$ and 4, these calculations will take some time to complete.
Empirically we find that the costs of numerical computations increase very rapidly with $N$, scaling $\propto\!N^5$.

Fortunately, because all $\cN = 4$ SYM fields transform in the adjoint representation, deviations from large-$N$ predictions are suppressed by two powers of $N$; they go as $1 / N^2$.
Thus even $N = 4$ should suffice to access the large-$N$ regime up to few-percent effects that may be comparable to our initial statistical uncertainties.
Systematically studying all of $N = 2$, 3 and 4 will also allow us to extrapolate $1 / N^2 \to 0$ as a further check of the approach to the large-$N$ limit.

\begin{figure*}[btp]
  \includegraphics[height=\figheight]{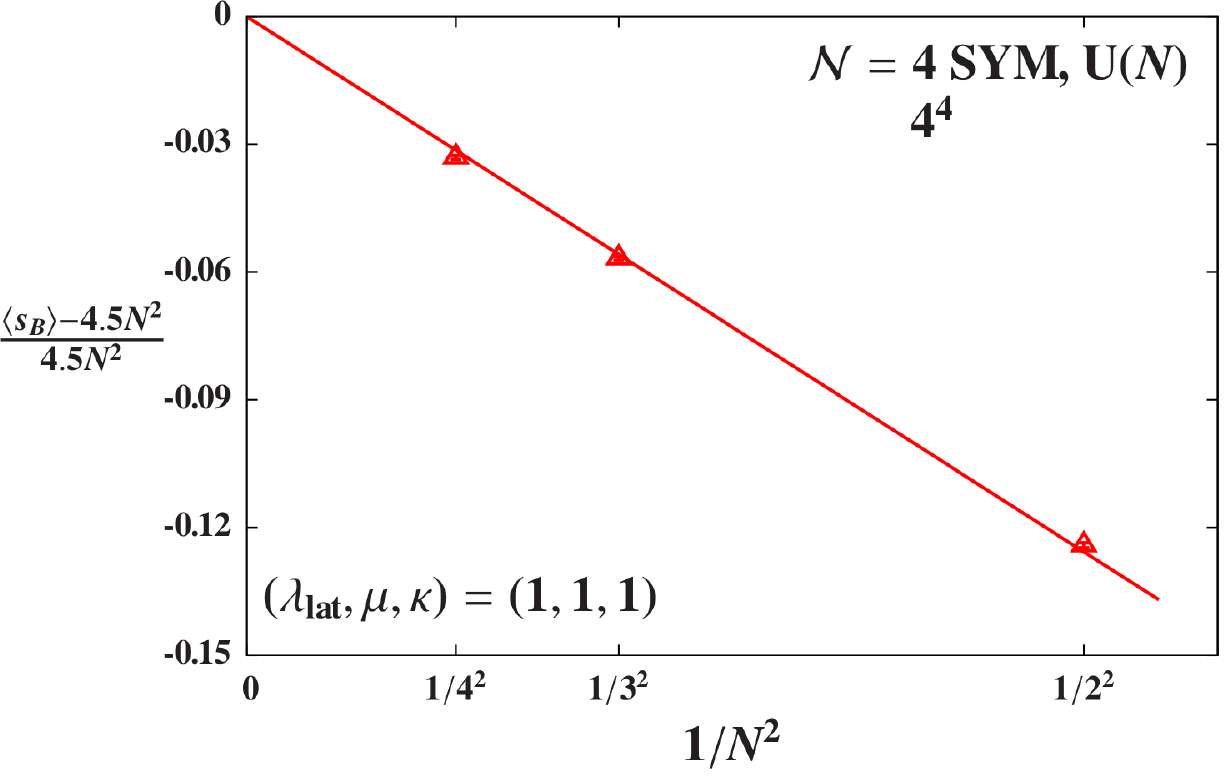}
  \caption{\label{fig:sB_N} Deviations of the bosonic action $\vev{s_B}$ from its exact supersymmetric value $9N^2 / 2$, as in \protect\fig{fig:sB}.  Results from $4^4$ ensembles with $(\lambda_{\rm lat}, \mu, \kappa) = (1, 1, 1)$ are plotted vs.\ $1 / N^2$ for gauge groups U(2), U(3) and U(4).  The deviations clearly scale $\propto\!1 / N^2$.  The red line is a linear fit constrained to vanish in the large-$N$ limit.}
\end{figure*}

\begin{figure*}[btp]
  \includegraphics[height=\figheight]{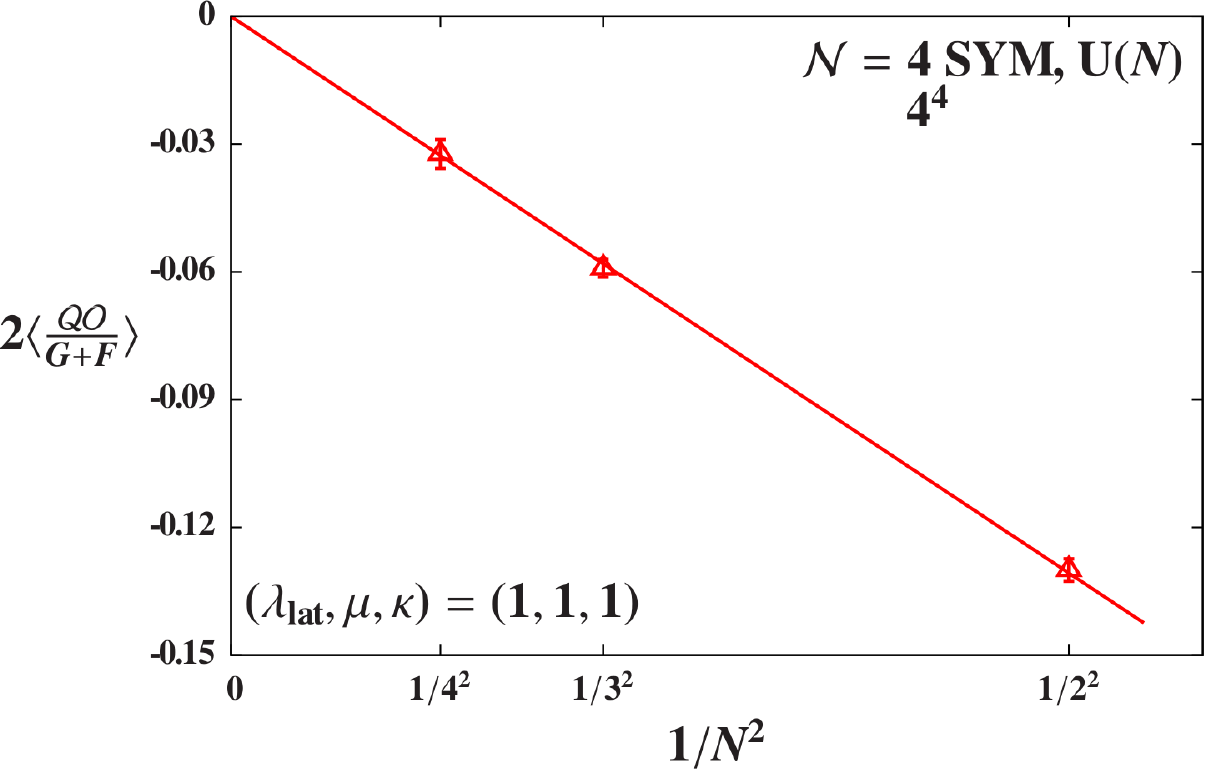}
  \caption{\label{fig:susy_N} Violations of the Ward identity $\vev{\cQ \cO} = 0$ as defined by \protect\eq{eq:susy_rel}, as in \protect\fig{fig:susy_breaking}.  Results from $4^4$ ensembles with $(\lambda_{\rm lat}, \mu, \kappa) = (1, 1, 1)$ are plotted vs.\ $1 / N^2$ for gauge groups U(2), U(3) and U(4).  The Ward identity violations clearly scale $\propto\!1 / N^2$.  The red line is a linear fit constrained to vanish in the large-$N$ limit.}
\end{figure*}

As a first example, in Figs.~\ref{fig:sB_N} and \ref{fig:susy_N} we show how our measures of supersymmetry breaking -- deviations of the bosonic action $\vev{s_B}$ from its exact supersymmetric value $9N^2 / 2$, and violations of the Ward identity $\vev{\cQ \cO} = 0$ as defined by \protect\eq{eq:susy_rel}, respectively -- depend on the gauge group.
We consider $N = 2$, 3 and 4 on $4^4$ lattices with fixed $(\lambda_{\rm lat}, \mu, \kappa) = (1, 1, 1)$.
Recall from Table~\ref{susy_vol_scaling} that even though the small volume contributes to supersymmetry breaking, this is only a percent-level effect for $4^4$ lattices, at least for $N = 2$.
We find that both observables clearly scale $\propto\!1 / N^2$.
That is, supersymmetry breaking is significantly suppressed as $N$ increases.
The straight lines in these plots are fits of the data to the simple form $A / N^2$; we find the slope $A \approx -0.5$.
This is encouraging evidence that our lattice theory continues to simulate $\cN = 4$ SYM to a good approximation in the theoretically interesting large-$N$ limit.
Indeed, the quality of the simulations improves as $N$ increases, at the price of significant increases in computational costs.


It is also important to investigate the pfaffian phase for U($N$) gauge groups with larger $N$, to ensure that U(2) is not simply a special case where the pfaffian is especially well behaved.
Of course, these computations are even more challenging than those discussed above, and at present we have only measured the U(3) and U(4) pfaffian phase for a handful of ensembles.
Table~\ref{pfaffian_N} shows our results for fixed $(\lambda_{\rm lat}, \mu, \kappa) = (1, 1, 1)$.
On $2^3\times4$ and $3^3\times4$ lattices, $\alpha$ essentially agrees within uncertainties for all of $N = 2$, 3 and 4.
The alternation of the overall sign with $N$ is irrelevant.
This alternating sign is also present in the matrix model obtained by dimensional reduction to zero dimensions, suggesting that it simply counts the number of generators.

\begin{table}[t]
  \caption{\label{pfaffian_N} The real part of the pfaffian phase, $\cos\alpha$, for $N = 2$, 3 and 4 on lattice volumes $2^3\times4$ and $3^3\times4$ with fixed $(\lambda_{\rm lat}, \mu, \kappa) = (1, 1, 1)$.  The phase remains small, and agrees within uncertainties for all gauge groups.  The irrelevant overall sign simply counts the number of U($N$) generators.}
  \centering
  \renewcommand\arraystretch{1.2}  
  \begin{tabular}{|c|c|c|c|}
    \hline
    ~Volume~      &  U(2)         &  U(3)             &  U(4)         \\\hline
    $2^3\times4$  &  0.99978(4)   &  $-$0.99980(3)    & ~0.99989(4)~  \\
    $3^3\times4$  & ~0.99914(22)~ & ~$-$0.99922(20)~  &  ---          \\
    \hline
  \end{tabular}
\end{table}

\section{\label{sec:potential}The potential}
In this section we describe our measurements of the potential between static test charges (in the fundamental representation), as determined from the behavior of Wilson loops.
The analysis has several parts.
In the next subsection we describe the general methodology we use to compute Wilson loops at many separations on the hypercubic representation of the $A_4^*$ lattice, and summarize the three different types of Wilson loops that we consider.
Section~\ref{sec:results} presents our results for the potential from $8^3\times 24$ lattice ensembles.

We provided a first look at the static potential in Ref.~\cite{Catterall:2012yq}.
That analysis considered only $8^4$ lattices, which prevented us from extending the Wilson loops to large time extents.
The only Wilson loops considered were those oriented along the axes of the lattice, which limited the number of spatial separations $r \equiv |\vec r|$ at which we had data points for $V(r)$.
And we did not correctly translate the fields and couplings of the lattice action defined on a hypercube, to the lattice action defined on the $A_4^*$ lattice, to the continuum theory.
In this work we address all of these issues.

\subsection{\label{sec:kinds}Lattice observables}
We extract the static potential from the asymptotic behavior of Wilson loops, including loops that are not oriented along the principal axes of the lattice.
This is awkward to do when explicitly constructing the Wilson loops as paths in the lattice.
Instead, we exploit a trick from QCD simulation technology, to measure loops with all available spatial separations $\vec r$.
The method is to gauge fix to Coulomb gauge, and then consider
\beq
  \label{eq:loop}
  W(\vec r, t) = \Tr P(\vec x, t, t_0) P^{\dag}(\vec x + \vec r, t, t_0),
\eeq
where $P(\vec x, t, t_0)$ is a product of temporal links at spatial location $\vec x$ extending from time-slice $t_0$ to time-slice $t_0 + t$.

We are imagining a transfer matrix carrying information from each time-slice to the next, and gauge fix to Coulomb gauge to ensure that this transfer matrix carries only gauge-singlet information.
Coulomb gauge fixing maximizes $\sum_i \Tr \cU_i$, where the sum runs over all the directions along which we are not propagating.
The technical issue for us is how to do the gauge fixing on the $A_4^*$ lattice.
Our anti-periodic boundary conditions identify the fourth link as the temporal direction, but none of the basis vectors in \eq{eq:basis} are orthogonal to $\hatbe_4$.
Since the first three links span each time-slice, we define Coulomb gauge by maximizing $\sum_i \Tr \cU_i$ for $i = 1$, 2 and 3 only.
In the future it may be worthwhile to explore whether including the fifth link in the Coulomb gauge condition would have any effect.

After gauge fixing we compute $W(r, t)$ with $r$ calculated using the basis vectors (\ref{eq:basis}) of the $A_4^*$ lattice (see Section~\ref{sec:A4star} for details).
From these data we extract the potential by performing a fit to $W(r, n_t) = \exp(-V(r)(a n_t))$, where ``$a$'' is the lattice spacing and $n_t$ is dimensionless.
This fit form corresponds to the standard $r$ dependence of the rectangular $r \times T$ Wilson loop in the continuum, when the temporal extent $T$ is much greater than $r$:
\begin{align}
  -\log W(r, T) & = T\times V(r) \\
  V(r) & = C g^2 N \int \frac{d^3 \vec k}{(2\pi)^3} \exp(i\vec k \cdot \vec r) D(k)
\end{align}
where $k = (0, \vec k)$ is enforced by the limit $T \gg r$.
For us $D(k) = D_{00}(k) = \vev{\cA_0(-k) \cA_0(k)}$ is the temporal component of the bosonic two-point function, in the language of Ref.~\cite{Catterall:2011pd}.
$C g^2 N$ is a Coulomb coupling constant, which to leading order in perturbation theory is $g^2 N / (4\pi)$ for supersymmetric Wilson loops.

On the lattice, $\vec r$ is defined through a set of integers $n_i$.
We translate our data for the potential to physical $\vec r = \sum_i n_i \hatbe_i$ using the three-dimensional reduction of the $A_4^*$ basis vectors (\ref{eq:basis}),
\begin{align}
  \hatbe_1 & = \left(\frac{1}{\sqrt{2}}, \frac{1}{\sqrt{6}}, \frac{1}{\sqrt{12}}\right)   &
  \hatbe_2 & = \left(-\frac{1}{\sqrt{2}}, \frac{1}{\sqrt{6}}, \frac{1}{\sqrt{12}}\right)  &
  \hatbe_3 & = \left(0, -\frac{2}{\sqrt{6}}, \frac{1}{\sqrt{12}}\right). \label{eq:3dbasis}
\end{align}
We expect the data to reflect the discrete symmetries of the $A_4^*$ lattice.
For example, the potential at $n = (1, 0, 0)$ and $n = (1, 1, 1)$ should be equal, since both correspond to $r = \sqrt{3 / 4}$.
This result can be checked by computing the potential using the propagator as defined in Ref.~\cite{Catterall:2011pd}.
Although we do not pursue this investigation in the present work, a direct comparison of our numerical data for different $\vec r$ would allow us to test the rotational invariance of our lattice system.

Unlike the results in previous sections, our analysis of the potential involves a direct comparison to the continuum expectations discussed above.
This requires us to account for the relation between the lattice and continuum couplings from \eq{eq:g_norm}, $\lambda_{\rm lat} = g^2 N \sqrt 5$ for the $A_4^*$ lattice.
As we are interested in the continuum potential, we will plot $V(r)$ as a function of the continuum 't~Hooft coupling $g^2 N = \lambda_{\rm lat} / \sqrt{5}$.

Finally, we measure three different kinds of Wilson loops.
The first is the ``usual'' Wilson loop formed from the complexified gauge links $\cU_a \in \gltwo$.
Next, we would like to probe the SU(2) sector of the theory, which corresponds to the projection to \sltwo discussed in Section~\ref{sec:det}.
In this context, we measure the ``determinant-divided'' Wilson loop,
\begin{equation}
  W_D(\vec r, t) \equiv \frac{W(\vec r, t)}{\det^{1 / N} W(\vec r, t)},
\end{equation}
with $N = 2$.
If this division removes the U(1) sector, we expect the Coulomb term will be reduced by a factor $(N^2 - 1) / N^2 = 3 / 4$.
Finally, we would like to remove the contribution of the scalars.
We can do that with the ``polar-projected'' loop: we define a unitary matrix $u$ corresponding to each link variable via a polar decomposition
\beq
  \cU = u H,
\eeq
where $H$ is hermitian, positive definite.
We expect loops $W_{pol}$ built from these unitary matrices $u$ to produce a Coulomb coefficient half of that corresponding to the usual Wilson loops, $C_{pol} = C / 2 = \frac{1}{2} \lambda_{\rm lat} / (4\pi \sqrt{5})$.

\subsection{\label{sec:results}Results}
The results presented here are from $8^3\times 24$ lattices.
While we have some ensembles with larger spatial volumes ($12^3$ and $16^3$), the size of the Coulomb coefficient is such that the signal disappears into the background noise by $r \approx 3$.
Larger volumes only help by giving more lattice sites over which we average the correlators (\ref{eq:loop}).
We chose $N_t = 24$ because Wilson loops can only be measured up to halfway across the lattice without awkward considerations about boundary conditions.
(The Wilson loop has no special periodicity properties.)
As we will see, our signals are not asymptotic until $t_{min} \approx 5$, so $N_t = 24$ (as opposed to the $N_t = 8$ considered in Ref.~\cite{Catterall:2012yq}) provides more data points below $t_{max} = N_t / 2$, making it easier to judge the quality of fits.

\begin{figure*}[btp]
  \includegraphics[height=\figheight]{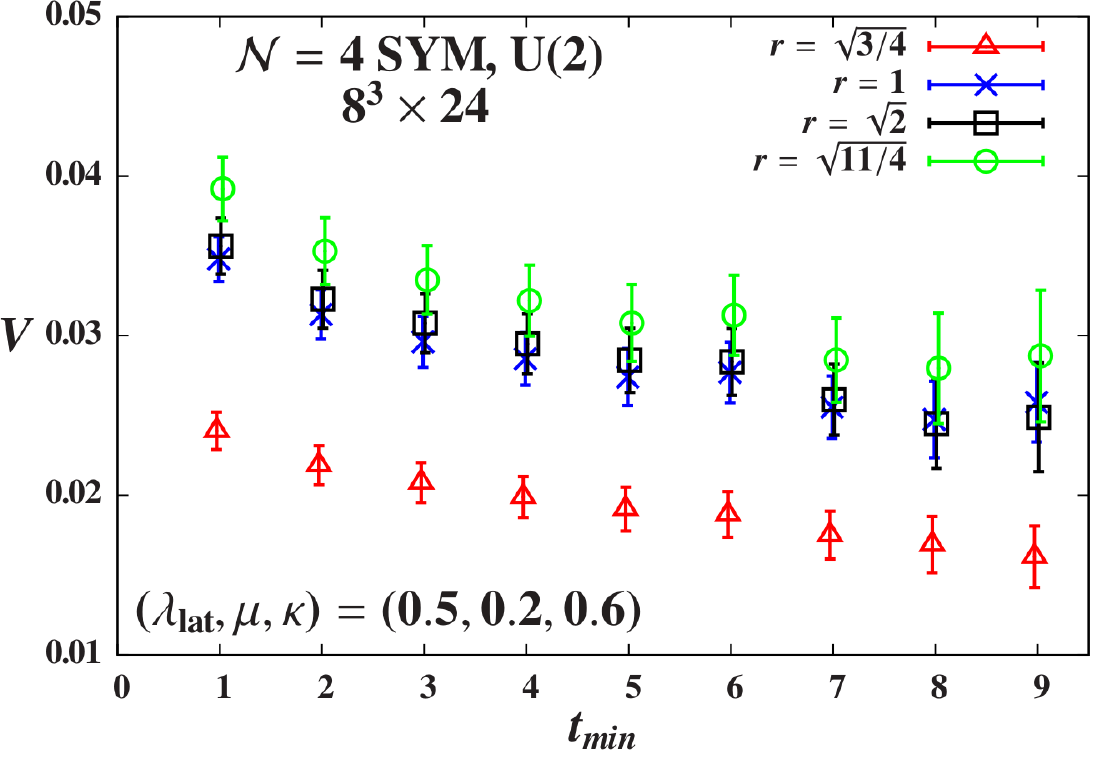}
  \caption{\label{fig:range0.5pot}To extract the static potential $V(r)$ we must determine the appropriate range $t_{min} \leq t < N_t / 2$ over which to fit the Coulomb gauge fixed Wilson loops to $W(r, t) = \exp(-V(r)t)$.  We do this by searching for a plateau in each $V(r)$ plotted vs.\ $t_{min}$.  These representative results for the $8^3\times24$ lattice ensemble with $(\lambda_{\rm lat}, \mu, \kappa) = (0.5, 0.2, 0.6)$ show plateaus appearing to begin at $t_{min} \approx 5$ for the four smallest $r = \sqrt{3 / 4}$, $1$, $\sqrt{2}$ and $\sqrt{11 / 4}$ on the $A_4^*$ lattice.  Other data sets are similar.}
\end{figure*}

The extraction of a potential begins with the measurement of Coulomb gauge fixed Wilson loops $W(\vec r, t)$.
To reliably fit $W(r, t) = \exp(-V(r)t)$, we must check that we obtain a stable $V(r)$ for each $r$.
To do this, we perform fits over different ranges of $t_{min} \leq t < N_t / 2$, and look for plateaus as we vary $t_{min}$.
Representative examples of such tests are shown in \fig{fig:range0.5pot}, considering the four smallest $r = \sqrt{3 / 4}$, $1$, $\sqrt{2}$ and $\sqrt{11 / 4}$ on the $A_4^*$ lattice, for the $8^3\times24$ lattice ensemble with $(\lambda_{\rm lat}, \mu, \kappa) = (0.5, 0.2, 0.6)$.
For each $r$ plateaus appear to begin at $t_{min} \approx 5$.
Other data sets are similar, and we conclude that fits beginning at $t_{min} = 5$--6 are safely asymptotic.

\begin{figure*}[btp]
  \includegraphics[height=\figheight]{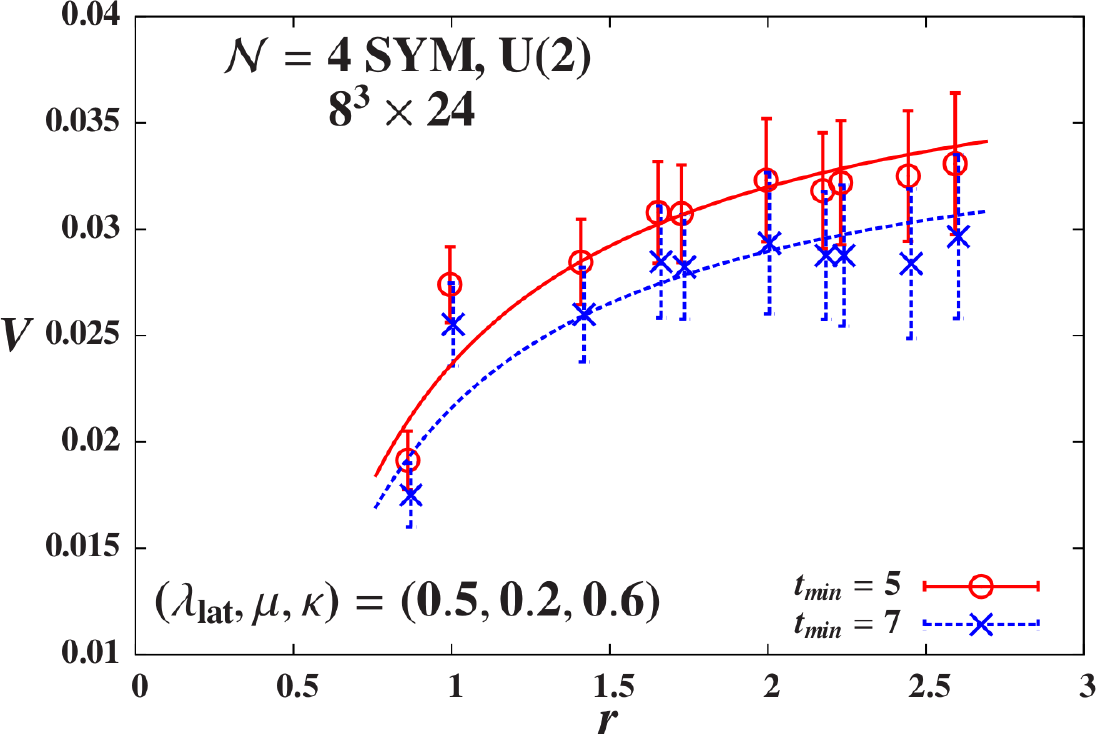}
  \caption{\label{fig:vr0.5pot}Representative results for the static potential $V(r)$ vs.\ $r$ from the usual Wilson loops, for the $(\lambda_{\rm lat}, \mu, \kappa) = (0.5, 0.2, 0.6)$ ensemble.  The curves are fits of $V(r)$ to the Coulombic form of \protect\eq{eq:coulomb}, for different $t_{min} = 5$ and 7 in the earlier fits to the Coulomb gauge fixed $W(r, t)$.}
\end{figure*}

\begin{figure*}[btp]
  \includegraphics[height=\figheight]{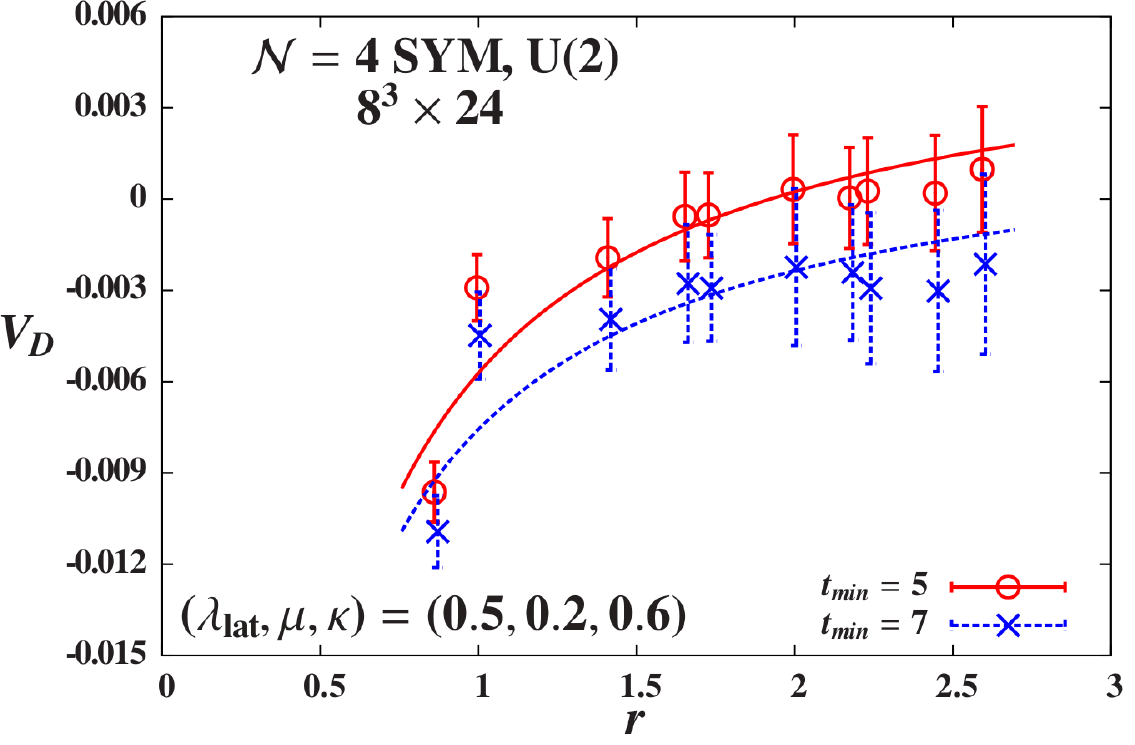}
  \caption{\label{fig:vrdet0.5pot}Representative results for the static potential $V_D(r)$ vs.\ $r$ from the determinant-divided Wilson loops, for the $(\lambda_{\rm lat}, \mu, \kappa) = (0.5, 0.2, 0.6)$ ensemble.  The curves are fits of $V_D(r)$ to the Coulombic form of \protect\eq{eq:coulomb}, for different $t_{min} = 5$ and 7 in the earlier fits to the Coulomb gauge fixed $W_D(r, t)$.}
\end{figure*}

\begin{figure*}[btp]
  \includegraphics[height=\figheight]{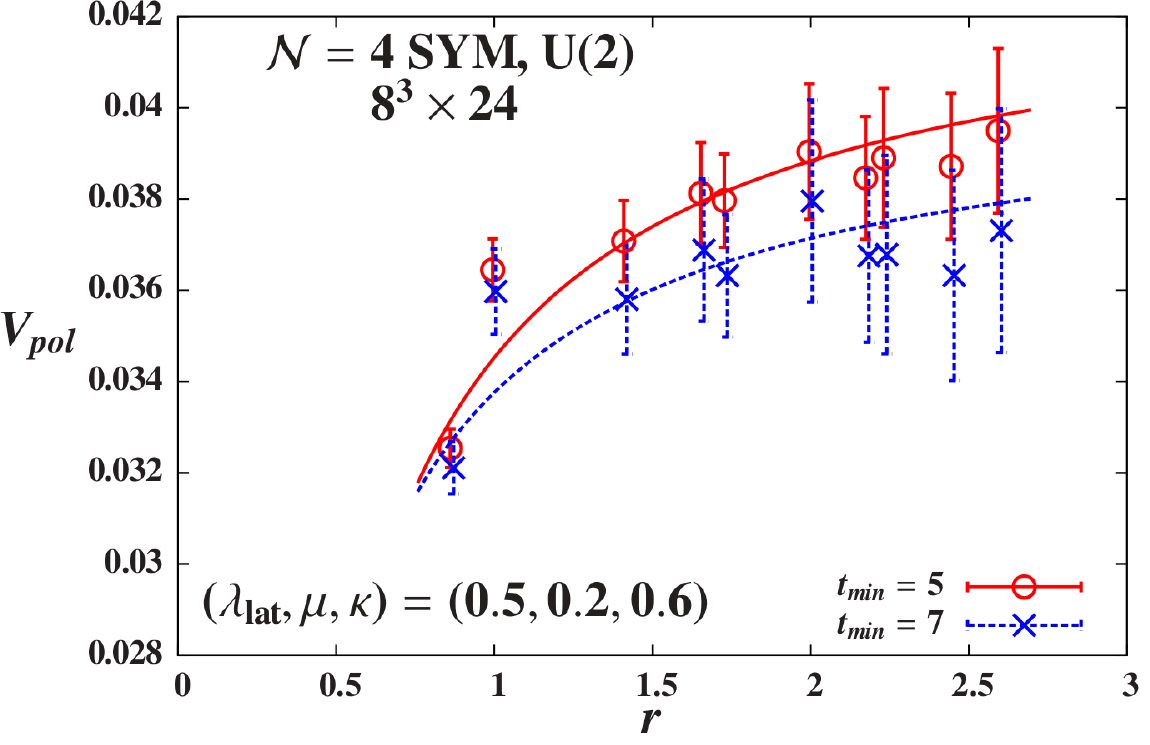}
  \caption{\label{fig:vrpol0.5pot}Representative results for the static potential $V_{pol}(r)$ vs.\ $r$ from the polar-projected Wilson loops, for the $(\lambda_{\rm lat}, \mu, \kappa) = (0.5, 0.2, 0.6)$ ensemble.  The curves are fits of $V_{pol}(r)$ to the Coulombic form of \protect\eq{eq:coulomb}, for different $t_{min} = 5$ and 7 in the earlier fits to the Coulomb gauge fixed $W_{pol}(r, t)$.}
\end{figure*}

We then fit the results for $V(r)$ to a Coulombic form,
\beq
  \label{eq:coulomb}
  V(r) = A - \frac{C}{r},
\eeq
and to a confining form
\beq
  \label{eq:confining}
  V(r) = A - \frac{C}{r} + \sigma r.
\eeq
Representative Coulomb potential fits are shown in \fig{fig:vr0.5pot}, again for the $(\lambda_{\rm lat}, \mu, \kappa) = (0.5, 0.2, 0.6)$ lattice ensemble, for two different $t_{min} = 5$ and 7.
This figure considers the potential from the usual Wilson loops; the corresponding results for the determinant-divided loops and the polar-projected loops are presented in Figs.~\ref{fig:vrdet0.5pot} and \ref{fig:vrpol0.5pot}, respectively.
The fits in each figure use all points shown, which span the range $\sqrt{3 / 4} \leq r \leq 2.6$. 
The upper bound is set by the maximum length that doesn't wrap around the $A_4^*$ lattice with $L = 8$.
As in our earlier work~\cite{Catterall:2012yq}, we do not find any statistically non-zero string tension $\sigma$ from fits to the confining form (\ref{eq:confining}).
This is consistent with the non-zero Polyakov loop in \fig{fig:pl824}.


\begin{figure*}[btp]
  \includegraphics[height=\figheight]{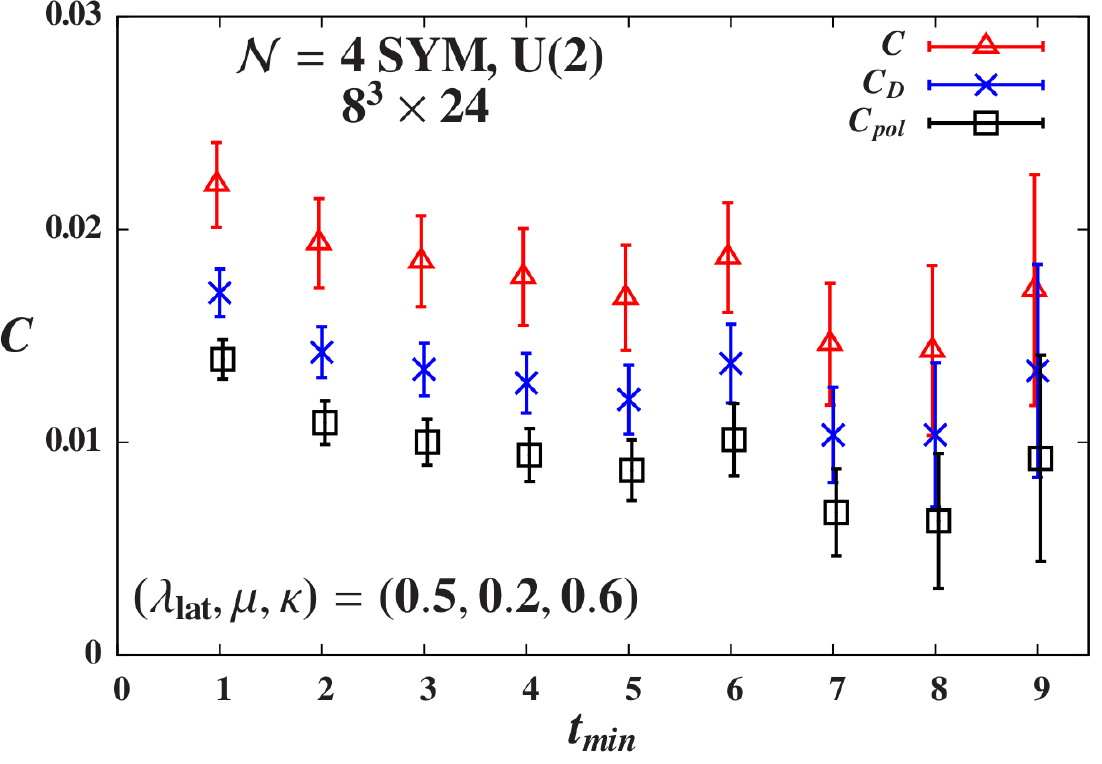}
  \caption{\label{fig:pot_tmin}Representative results for the static potential Coulomb coefficients $C$ vs.\ the $t_{min}$ used in the fits to Coulomb gauge fixed Wilson loop data, for the $(\lambda_{\rm lat}, \mu, \kappa) = (0.5, 0.2, 0.6)$ ensemble.  The three $C$, $C_D$ and $C_{pol}$ correspond to the usual Wilson loops, determinant-divided loops and polar-projected loops, respectively.  Plateaus in the Coulomb coefficients extend over a wide range of $t_{min}$, and the other data sets behave similarly.}
\end{figure*}

In Figs.~\ref{fig:vr0.5pot}, \ref{fig:vrdet0.5pot} and \ref{fig:vrpol0.5pot} we can see that the fit results for the Coulomb coefficients $C$ may depend on the chosen value of $t_{min}$, in addition to the value of $\lambda_{\rm lat}$.
Therefore we should also monitor $C$ itself as we vary $t_{min}$, in the same way we considered $V(r)$ in \fig{fig:range0.5pot}.
Representative results for $C$, $C_D$ and $C_{pol}$ vs.\ $t_{min}$ are shown in \fig{fig:pot_tmin}, again for the $(\lambda_{\rm lat}, \mu, \kappa) = (0.5, 0.2, 0.6)$ lattice ensemble.
We consistently see plateaus in the Coulomb coefficients across a wide range of $t_{min}$, although the signal degrades at larger $t_{min}$ and for larger $\lambda_{\rm lat}$.
The strongest coupling with a reliable signal is $\lambda_{\rm lat} = 4$.

\begin{figure*}[btp]
  \includegraphics[height=\figheight]{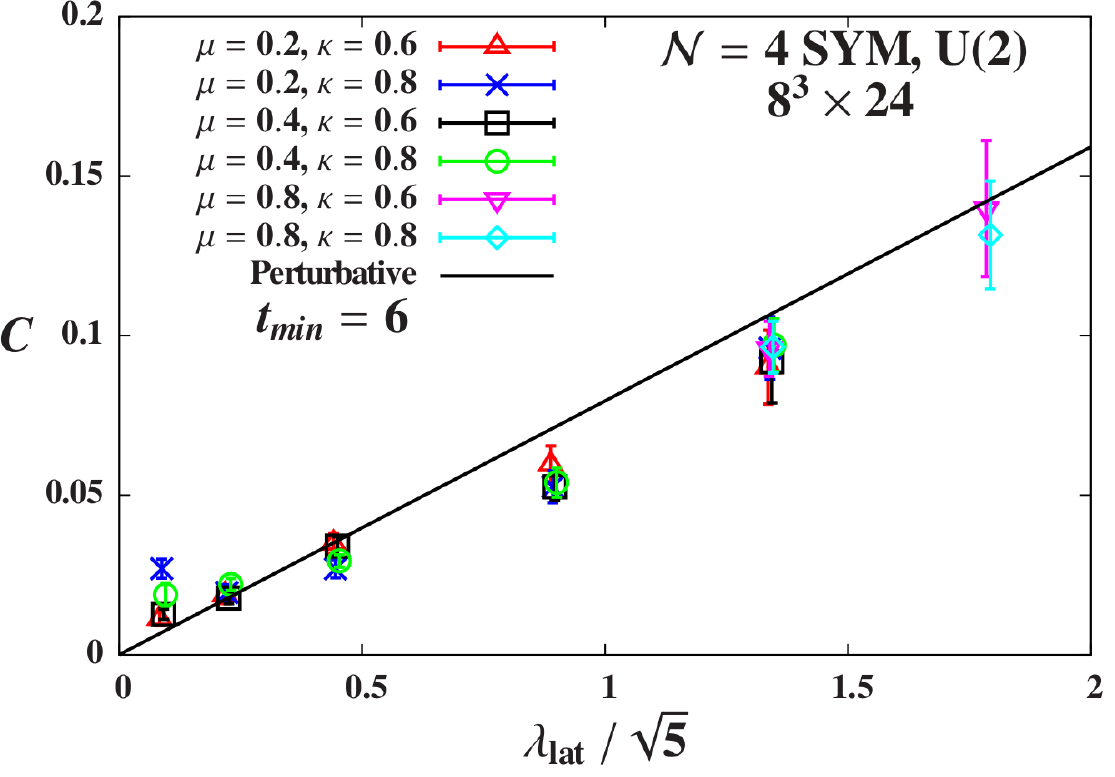}
  \caption{\label{fig:cvl}Static potential Coulomb coefficients $C$ for the usual Wilson loops, based on fits with $t_{min} = 6$ and $r \leq 2.6$.  Six $8^3\times24$ data sets with $(\mu, \kappa) = (0.2, 0.6)$, (0.2, 0.8), (0.4, 0.6), (0.4, 0.8), (0.8, 0.6) and (0.8, 0.8) are plotted vs.\ the continuum 't~Hooft coupling $\lambda_{\rm lat} / \sqrt{5}$.  No significant dependence on $\mu$ or $\kappa$ is visible.  The line is the leading-order perturbative prediction from \protect\eq{eq:Cperturb}, not a fit.  The results agree quite well with this perturbative prediction, and show no sign of the $C \propto \sqrt{\lambda}$ scaling predicted at strong coupling in the large-$N$ limit.}
\end{figure*}

\begin{figure*}[btp]
  \includegraphics[height=\figheight]{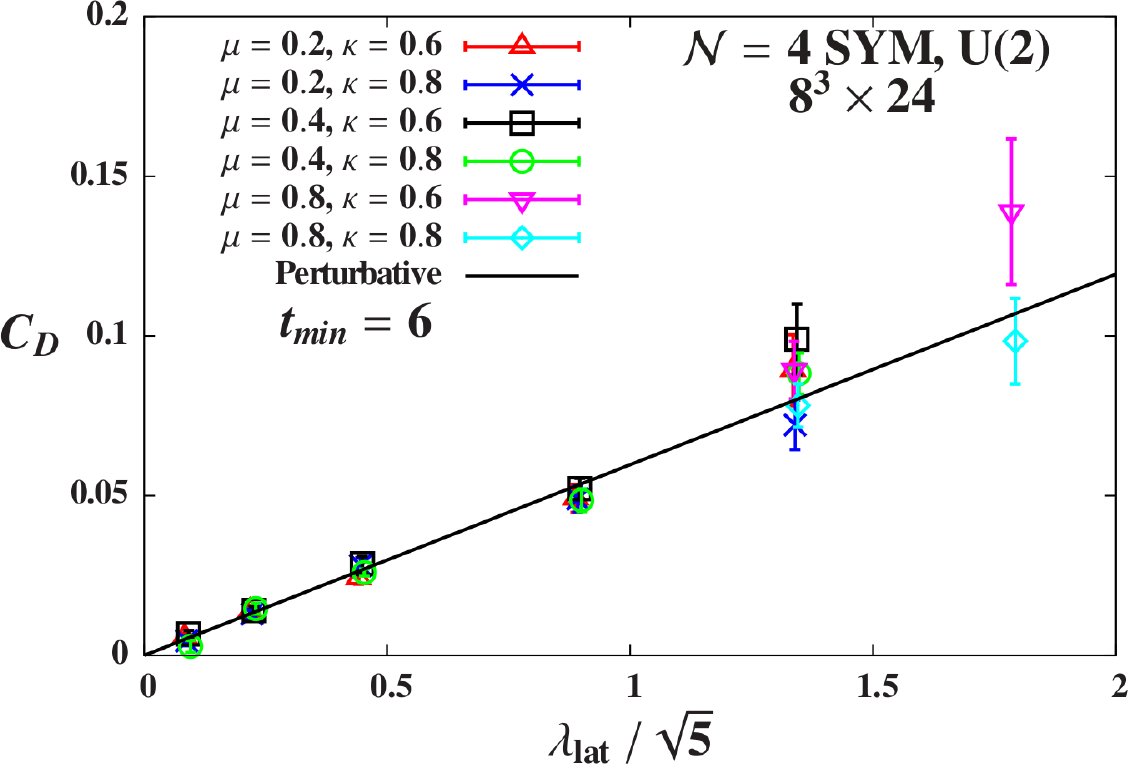}
  \caption{\label{fig:cdetvl}Static potential Coulomb coefficients $C_D$ for the determinant-divided Wilson loops, based on fits with $t_{min} = 6$ and $r \leq 2.6$.  Six $8^3\times24$ data sets with $(\mu, \kappa) = (0.2, 0.6)$, (0.2, 0.8), (0.4, 0.6), (0.4, 0.8), (0.8, 0.6) and (0.8, 0.8) are plotted vs.\ the continuum 't~Hooft coupling $\lambda_{\rm lat} / \sqrt{5}$.  No significant dependence on $\mu$ or $\kappa$ is visible.  The line is the leading-order perturbative prediction from \protect\eq{eq:Cperturb}, not a fit.  The results agree quite well with this perturbative prediction, and show no sign of the $C_D \propto \sqrt{\lambda}$ scaling predicted at strong coupling in the large-$N$ limit.  The axes cover the same range as those in \protect\fig{fig:cvl}.}
\end{figure*}

\begin{figure*}[btp]
  \includegraphics[height=\figheight]{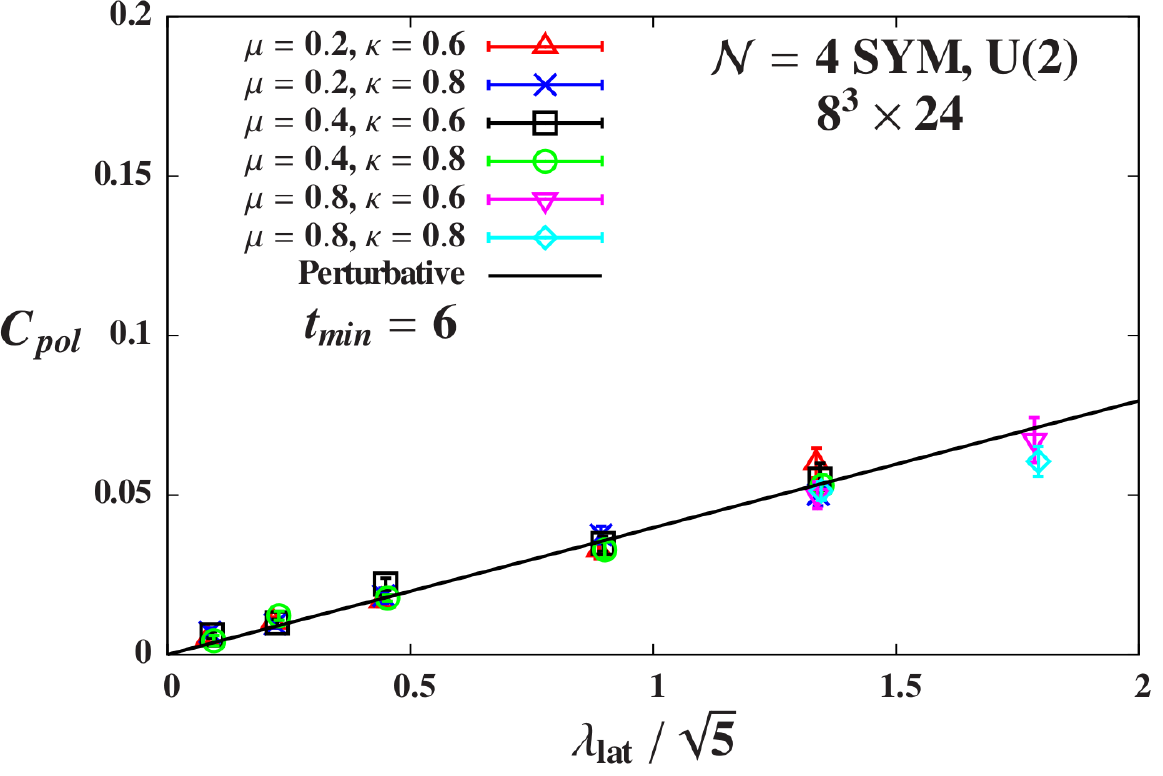}
  \caption{\label{fig:cpolvl}Static potential Coulomb coefficients $C_{pol}$ for the polar-projected Wilson loops, based on fits with $t_{min} = 6$ and $r \leq 2.6$.  Six $8^3\times24$ data sets with $(\mu, \kappa) = (0.2, 0.6)$, (0.2, 0.8), (0.4, 0.6), (0.4, 0.8), (0.8, 0.6) and (0.8, 0.8) are plotted vs.\ the continuum 't~Hooft coupling $\lambda_{\rm lat} / \sqrt{5}$.  No significant dependence on $\mu$ or $\kappa$ is visible.  The line is the leading-order perturbative prediction from \protect\eq{eq:Cperturb}, not a fit.  The results agree quite well with this perturbative prediction, and show no sign of the $C_{pol} \propto \sqrt{\lambda}$ scaling predicted at strong coupling in the large-$N$ limit.  The axes cover the same range as those in \protect\fig{fig:cvl}.}
\end{figure*}

We can now present our results for the Coulomb coefficients as functions of the continuum 't~Hooft coupling $\lambda_{\rm lat} / \sqrt{5}$.
\fig{fig:cvl} shows $C$ for the static potential from the usual Wilson loops, based on fits using $t_{min} = 6$ and $r \leq 2.6$.
The corresponding results for the determinant-divided loops and the polar-projected loops are presented in Figs.~\ref{fig:cdetvl} and \ref{fig:cpolvl}, respectively.
The lines are the naive predictions of lowest-order perturbation theory combined with the scaling expectations from Section~\ref{sec:kinds}:
\begin{align}
  \label{eq:Cperturb}
  C & = \frac{\lambda_{\rm lat} / \sqrt{5}}{4\pi} &
  C_D & = \frac{3}{4} \frac{\lambda_{\rm lat} / \sqrt{5}}{4\pi} &
  C_{pol} & = \frac{1}{2} \frac{\lambda_{\rm lat} / \sqrt{5}}{4\pi}
\end{align}
for the usual, determinant-divided, and polar-projected Wilson loops, respectively.


\begin{figure*}[btp]
  \includegraphics[height=\figheight]{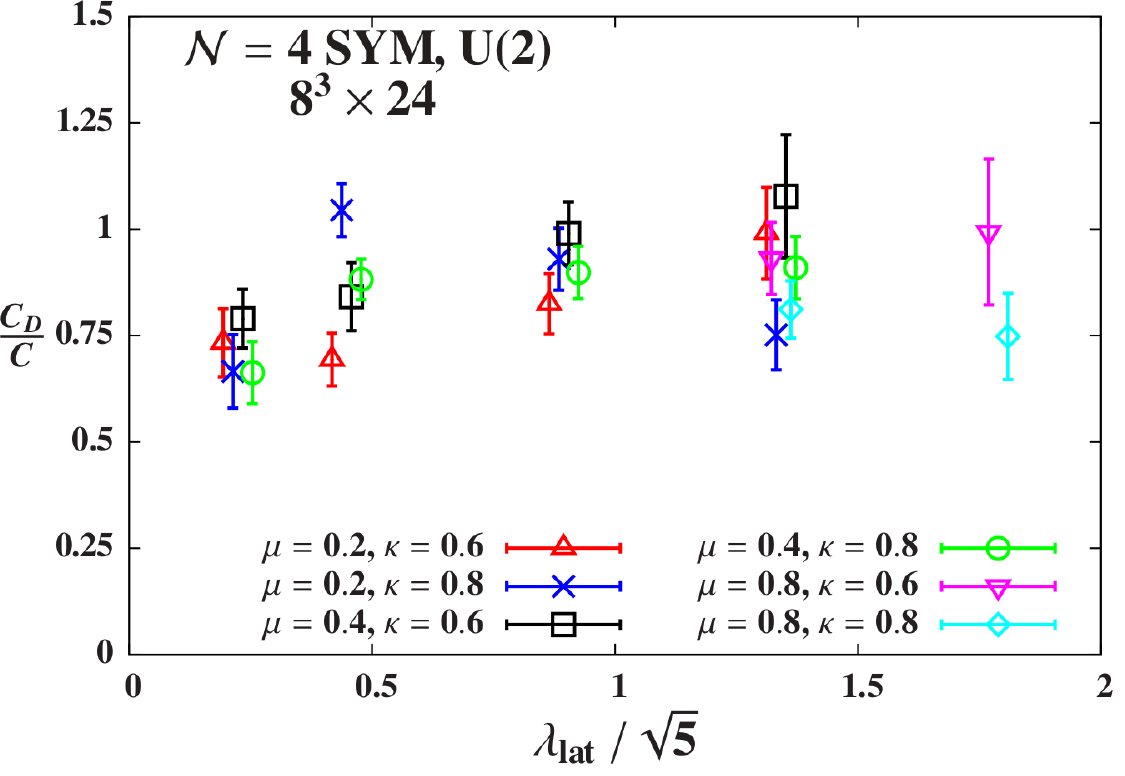}
  \caption{\label{fig:ratdet}Ratios $C_D / C$ of static potential Coulomb coefficients for the determinant-divided Wilson loops relative to those for the usual loops, based on fits with $t_{min} = 6$ and $r \leq 2.6$.  Six $8^3\times24$ data sets with $(\mu, \kappa) = (0.2, 0.6)$, (0.2, 0.8), (0.4, 0.6), (0.4, 0.8), (0.8, 0.6) and (0.8, 0.8) are plotted vs.\ the continuum 't~Hooft coupling $\lambda_{\rm lat} / \sqrt{5}$.  No significant dependence on $\mu$ or $\kappa$ is visible.  The results, although noisy, are consistent with the expected ratio of $3 / 4$ from \protect\eq{eq:Cperturb}.}
\end{figure*}

\begin{figure*}[btp]
  \includegraphics[height=\figheight]{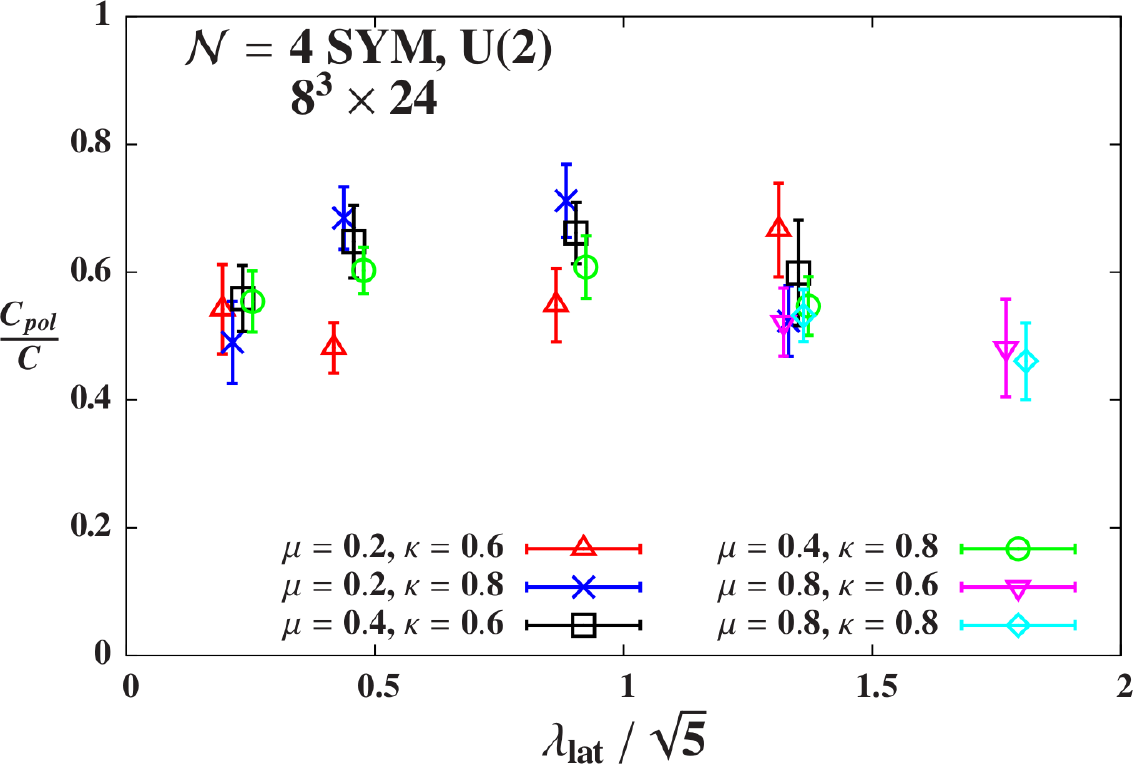}
  \caption{\label{fig:ratpol}Ratios $C_{pol} / C$ of static potential Coulomb coefficients for the polar-projected Wilson loops relative to those for the usual loops, based on fits with $t_{min} = 6$ and $r \leq 2.6$.  Six $8^3\times24$ data sets with $(\mu, \kappa) = (0.2, 0.6)$, (0.2, 0.8), (0.4, 0.6), (0.4, 0.8), (0.8, 0.6) and (0.8, 0.8) are plotted vs.\ the continuum 't~Hooft coupling $\lambda_{\rm lat} / \sqrt{5}$.  No significant dependence on $\mu$ or $\kappa$ is visible.  The results, although noisy, are consistent with the expected ratio of $1 / 2$ from \protect\eq{eq:Cperturb}.}
\end{figure*}

The perturbative predictions describe our results quite well, which is not too surprising given the relatively small continuum 't~Hooft couplings $\lambda_{\rm lat} / \sqrt{5} < 1.8$ that we study.
There is a famous continuum prediction that $C \propto \sqrt{\lambda}$ at strong coupling in the large-$N$ limit (with $\lambda \ll N$)~\cite{Rey:1998ik, Maldacena:1998im}.
We see no sign of such behavior in Figs.~\ref{fig:cvl} through \ref{fig:cpolvl}, as we would expect based on the small $N = 2$ we currently consider.
It is not clear whether our future large-volume studies of the U(3) and U(4) systems will involve large enough $N$ to probe this predicted $\sqrt{\lambda}$ dependence.
In the meantime, we can more directly check our scaling expectations for the three different kinds of potentials by considering ratios of the different Coulomb coefficients for fixed $(\lambda_{\rm lat}, \mu, \kappa)$.
Results for $C_D / C$ are presented in \fig{fig:ratdet}, while $C_{pol} / C$ is shown in \fig{fig:ratpol}.
In each case we observe the expected value of $3 / 4$ or $1 / 2$, respectively, although the results are fairly noisy.


We expect that, as in lattice QCD simulations~\cite{Hasenfratz:2001hp}, the signal for the potential can be improved by employing ``fat'', or smeared, link variables.
We are currently exploring this possibility.

\section{Conclusions}
In this paper we have reported first results from large-scale lattice studies of $\cN = 4$ SYM based on the SU($N$) gauge group, focusing on the case $N = 2$.
We employ a lattice action that retains an exact supersymmetry for $(\mu, \kappa) = (0, 0)$, where $\mu$ is a bosonic mass parameter that regulates the flat directions, and $\kappa$ is the coupling in a new plaquette determinant term in the action, which enforces an approximate projection from U($N$) down to SU($N$).
When $\kappa = 0$ we observe a transition to a strongly-coupled lattice phase dominated by U(1) monopoles.
This is a pure lattice artifact, with no analog in the continuum theory where the U(1) sector decouples.
We can remove this lattice phase for arbitrarily large 't~Hooft coupling by setting $\kappa \geq 0.5$.

The existence of an exact supersymmetry $\cQ$ ensures that the couplings to all supersymmetry-breaking counterterms vanish in the limit $(\mu, \kappa) \to (0, 0)$.
In practice, we observe that the soft breaking of this supersymmetry is largely determined by $\kappa$, is at most $\cO(10\%)$ for bare 't~Hooft couplings $\lambda_{\rm lat} \lesssim 1$, and (on small lattice volumes) is suppressed $\sim 1 / N^2$.
Ref.~\cite{Catterall:2013roa} shows how the recovery of the other 15 supersymmetries in the $1 / L \to 0$ continuum limit follows from certain discrete R symmetries.
In this paper we carried out a first numerical study of these R symmetries at the scale of the plaquette.
The R symmetry breaking we observe is rather mild, $\cO(10\%)$, which encourages more thorough investigations that we will carry out in the near future.
This future work will focus on larger Wilson loops, to more directly probe the restoration of $\cN = 4$ supersymmetry in the long-distance effective theory, and thus in the continuum limit.

Next, we presented evidence that our lattice theory does not suffer from a genuine sign problem despite possessing a complex pfaffian.
Employing new parallel software to directly evaluate the pfaffian, we find that it is approximately real and positive on all lattice volumes we can explore.
Fluctuations in the phase show no significant dependence on either the lattice volume or the gauge group U($N$) for $N = 2$, 3 and 4.
Further lattice studies of the $\cN = 4$ pfaffian, hopefully leading to improved qualitative understanding of this issue, will be the focus of another future publication.

Finally, we reported results for the static potential on the $8^3\times 24$ lattice ensembles listed in Table~\ref{data_sets}, significantly improving the initial study of Ref.~\cite{Catterall:2012yq}.
We confirm the basic conclusion of that work, finding that the potential appears Coulombic at both weak and strong 't~Hooft coupling.
Furthermore, by more carefully relating our lattice calculations to the continuum theory, we observe good agreement with leading-order perturbative predictions for the Coulomb coefficients.
The Coulomb coefficients extracted from Wilson loops of the full complexified gauge links, from loops projected to $\slN$, and from loops of unitary matrices that omit the scalar-field contributions, show the expected relative magnitudes (\ref{eq:Cperturb}).

Obviously there are many interesting directions to be pursued by future lattice studies of $\cN = 4$ SYM.
These include the issues of the pfaffian and of supersymmetry restoration in the continuum limit discussed above.
We could also revisit the eigenvalue spectrum studied by Ref.~\cite{Weir:2013zua}, for our new large-volume lattice ensembles with non-zero $\kappa$.
In addition, our results for the static potential and its Coulomb coefficients (e.g., Figs.~\ref{fig:vr0.5pot}, \ref{fig:cvl} and \ref{fig:ratdet}) remain rather noisy.
We believe this situation can be improved by developing techniques to smear the (non-unitary) gauge links on the $A_4^*$ lattice.
Such smearing may be necessary to obtain precise results for other correlation functions of the theory, which we are also actively investigating.
In particular it would be exciting for lattice calculations to make predictions for the anomalous dimensions of single-trace operators like the Konishi~\cite{D'Hoker:2002aw}.
Such lattice results would be complementary to those obtainable in perturbation theory or via the conformal bootstrap program~\cite{Beem:2013qxa}.

Since continuum studies of $\cN = 4$ SYM tend to be anchored in the large-$N$ limit, it will also be important for us to investigate the U(3) and U(4) gauge groups on larger lattices than those considered in Section~\ref{sec:largeN}.
Specifically, we have started generating $8^3\times24$ lattice ensembles for both of these systems, calculations that will take some time to complete.
On these saved configurations, we can easily carry out the same studies as discussed above for gauge group U(2).
For example, we can analyze the static potential and explore whether the coupling dependence of its Coulomb coefficient approaches the famous prediction $C \propto \sqrt{\lambda}$~\cite{Rey:1998ik, Maldacena:1998im}.

Finally, there may be alternative formulations or implementations of lattice $\cN = 4$ SYM that would be advantageous to explore.
As one example, it might be possible to drop the supersymmetry-breaking bosonic mass term (\ref{eq:mass}) from our action, by instead using twisted temporal boundary conditions to regulate the SU($N$) flat directions.
Since the relevant plaquettes involve two links crossing the temporal boundary in opposite directions, we would need to impose twists that aren't elements of the center symmetry group. 
It remains unclear whether this approach would work.
It will be interesting to consider, and we are confident there are many further possibilities that have not yet been conceived.

\vspace{12 pt}
\noindent {\sc Acknowledgments:}~We thank A.~Veernala, C.~DeTar, D.~Weir, M.~Hanada, A.~Cherman and A.~Hasenfratz for helpful discussions and suggestions.
We are grateful for hospitality and support from the Aspen Center for Physics (U.S.~National Science Foundation grant PHYS-1066293) when this project was initiated.
This work was also supported by the U.S.~Department of Energy (DOE) under grants DE-SC0008669 and DE-SC0009998 (S.C., D.S.) and DE-SC0010005 (T.D.). 
Numerical calculations were carried out on the HEP-TH cluster at the University of Colorado and on the DOE-funded USQCD facilities at Fermilab.

\section*{Appendix: Data sets}
Table~\ref{data_sets} summarizes the large-volume U(2) lattice ensembles considered in this work, omitting the ensembles with $\kappa = 0$ that were only used to explore monopole condensation in Section~\ref{sec:UvsSU}.
We impose anti-periodic temporal boundary conditions for the fermionic (but not bosonic) fields.
We generate gauge configurations through the phase-quenched rational hybrid Monte Carlo algorithm discussed in Ref.~\cite{Catterall:2011cea}, using new parallel software that will be presented in a forthcoming publication~\cite{Schaich:2014}.
For each ensemble specified by $\left\{V, \lambda_{\rm lat}, \mu, \kappa\right\}$, Table~\ref{data_sets} lists the total number of molecular dynamics time units (MDTU) generated, our thermalization cut, the number of measurements after the thermalization cut, and the resulting number of jackknife blocks used in analyses.
All measurements are separated by 10~MDTU, and we use jackknife blocks with a fixed size of 100 MDTU (ten measurements).

\begin{table}[p]
  \caption{\label{data_sets} Large-volume lattice ensembles for $\cN = 4$ SYM with gauge group U(2).}
  \centering
  \renewcommand\arraystretch{1.2}  
  \begin{tabular}{|c|ccc||c|c|c|c|}
    \hline
    Volume          & $\lambda_{\rm lat}$ & $\mu$ & $\kappa$  & ~MDTU~  & ~Therm.~  & ~\# meas.~ & ~\# blocks~ \\\hline\hline
                    &           & 0.2     & 0.6       & 1500  & 400     & 110     & 11        \\
    $8^3\times24$   & 0.2       & 0.2     & 0.8       & 1300  & 200     & 110     & 11        \\
                    &           & 0.4     & 0.6       & 1700  & 300     & 140     & 14        \\
                    &           & 0.4     & 0.8       & 1300  & 300     & 100     & 10        \\
    \hline
                    &           & 0.2     & 0.6       & 1300  & 200     & 110     & 11        \\
    $8^3\times24$   & 0.5       & 0.2     & 0.8       & 1300  & 200     & 110     & 11        \\
                    &           & 0.4     & 0.6       & 2000  & 300     & 170     & 17        \\
                    &           & 0.4     & 0.8       & 2000  & 300     & 170     & 17        \\
    \hline
                    &           & 0.2     & 0.6       & 1500  & 500     & 100     & 10        \\
                    &           & 0.2     & 0.8       & 1300  & 200     & 110     & 11        \\
    $8^3\times24$   & 1.0       & 0.4     & 0.6       & 1600  & 500     & 110     & 11        \\
                    &           & 0.4     & 0.8       & 3000  & 300     & 270     & 27        \\
                    &           & 1.0     & 1.0       & 1605  & 405     & 120     & 12        \\
    \hline
                    &           & 0.2     & 0.6       & 1300  &  300    & 100     & 10        \\
    $8^3\times24$   & 2.0       & 0.2     & 0.8       & 1600  &  600    & 100     & 10        \\
                    &           & 0.4     & 0.6       & 2500  & 1400    & 110     & 11        \\
                    &           & 0.4     & 0.8       & 1300  &  300    & 100     & 10        \\
    \hline
                    &           & 0.2     & 0.6       & 1300  &  300    & 100     & 10        \\
                    &           & 0.2     & 0.8       & 1800  &  800    & 100     & 10        \\
    $8^3\times24$   & 3.0       & 0.4     & 0.6       & 1300  &  300    & 100     & 10        \\
                    &           & 0.4     & 0.8       & 2200  &  500    & 170     & 17        \\
                    &           & 0.8     & 0.6       & 2000  &  700    & 130     & 13        \\
                    &           & 0.8     & 0.8       & 2000  &  500    & 150     & 15        \\
    \hline
    $8^3\times24$   & 4.0       & 0.8     & 0.6       & 2000  &  600    & 140     & 14        \\
                    &           & 0.8     & 0.8       & 2200  &  700    & 150     & 15        \\
    \hline
    $8^3\times24$   & 5.0       & 0.8     & 0.6       & 1400  &  300    & 110     & 11        \\
                    &           & 0.8     & 0.8       & 1500  &  400    & 110     & 11        \\
    \hline
    $12^3\times24$  & 1.0       & 1.0     & 1.0       & 3000  & 2000    & 100     & 10        \\
    \hline
    $16^3\times32$  & 1.0       & 1.0     & 1.0       & 3500  & 2200    & 130     & 13        \\
    \hline
  \end{tabular}
\end{table}

\bibliographystyle{apsrev}
\bibliography{N4SUSYII}
\end{document}